\documentclass[preprint]{aastex}

 \def\baselinestretch{1.15}
\newcommand{\myemail}{pfyan0822@sina.com; yuanqirong@njnu.edu.cn}
\newcommand{\kms}{km~s$^{-1}$}

\shorttitle{ Multicolor Photometry of the Merging Cluster A2319 }

\shortauthors{Yan, Yuan, Zhang, Zhou, \& Jiang }

\begin{document}

\title{MULTICOLOR PHOTOMETRY OF THE MERGING GALAXY CLUSTER A2319: DYNAMICS AND STAR FORMATION PROPERTIES }


\author{\large Peng-Fei Yan\altaffilmark{1,2}, Qi-Rong Yuan\altaffilmark{1},
Li Zhang\altaffilmark{3},
and Xu Zhou\altaffilmark{4}
% , \and Jian-Sheng Chen\altaffilmark{4},
}

\altaffiltext{1}{Department of Physics and Institute of Theoretical Physics, 
		  Nanjing Normal University, Nanjing 210023, China; \myemail}
\altaffiltext{2}{School of Mathematics and Physics, Qingdao
                University of Science and Technology, Qingdao 266061, China;}
\altaffiltext{3}{QuFu Education Bureau, QuFu 273100, China;}    
\altaffiltext{4}{National Astronomical Observatories, Chinese
Academy of Sciences, Beijing 100012, China}
% zhouxu@vega.bac.pku.edu.cn

\begin{abstract}

Asymmetric X-ray emission and powerful cluster-scale radio halo
% and a very large velocity dispersion (\underline{$\sim 2001$ km s$^{-1}$})
indicate that A2319 is a merging cluster of galaxies. This paper presents our multicolor photometry for A2319 with 15 optical intermediate filters in the Beijing-Arizona-Taiwan-Connecticut (BATC) system. There are 142 galaxies with known spectroscopic redshifts within the viewing field of $58'\times58'$ centered on this rich cluster, including 128 member galaxies (called sample I).A large velocity dispersion in the rest frame, $1622^{+91}_{-70}$ \kms,  suggests a merger dynamics in A2319. The contour map of projected density and localized velocity structure confirm the so-called A2319B substructure, at $\sim 10'$ NW to the main concentration A2319A. The spectral energy distributions (SEDs) of more than 30,000 sources are obtained in our BATC photometry down to $V \sim 20$ mag. A $u$-band ($\sim$ 3551\AA ) image with better seeing and spatial resolution, obtained with the Bok 2.3m telescope at Kitt Peak, is taken to make star-galaxy separation and distinguish the overlapping contamination in the BATC aperture photometry. With color-color diagrams and photometric redshift technique, 233 galaxies brighter than $h_{\rm BATC}=19.0$ are newly selected as member candidates after an exclusion of false candidates with contaminated BATC SEDs by eyeballing checking the $u$-band Bok image. The early-type galaxies are found to follow a tight color-magnitude correlation.
Based on sample I and the enlarged sample of member galaxies (called sample II), subcluster A2319B is confirmed. The star formation properties of cluster galaxies are derived with the evolutionary synthesis model, PEGASE, assuming a Salpeter IMF and an exponentially decreasing star formation rate (SFR). A strong environmental effect on star formation histories is found in the manner that galaxies in the sparse regions have various star formation histories, while galaxies in the dense regions are found to have shorter SFR time scales, older stellar ages, and higher ISM metallicities. For the merging cluster A2319, local surface density is a better environmental indicator rather than the clustercentric distance. Compared with the well-relaxed cluster A2589, a higher fraction of star-forming galaxies is found in A2319, indicating that the galaxy-scale turbulence stimulated by the subcluster merger might have played a role in triggering the star formation activity.  

\end{abstract}


%% and \citet commands to identify citations. The citations are

\clearpage

\section{INTRODUCTION}

Galaxy clusters are the most massive gravitationally bound systems in the universe, and they have aroused more and more interests since Abell (1958) achieved the catalogue of 2712 rich galaxy clusters. According to the hierarchical scenario of structure formation, galaxy clusters grow by episodic mergers of smaller mass units (groups and poor clusters) and through the continuous accretion of gas from the surrounding intergalactic medium (IGM) (West et al.1991; Colberg et al. 2000). All massive clusters have experienced several mergers in their histories, and quantitative measurements of the substructures in nearby clusters (Geller \& Beers 1982; Dressler \& Shectman 1988; Mohr et al. 1995; Buote \& Tsai 1996) show that most present-day clusters are still in the process of accreting matter (Dressler \& Shectman 1988; O$^{'}$Hara et al. 2006). Improved capabilities (e.g., sensitivity, spectral and spatial resolution) of multi-wavelength telescopes allow us to investigate the formation and evolution of galaxy clusters in detail, which may shed light on the theories of structure formation and galaxy evolution (Kauffmann et al. 1999).

The rich cluster, A2319, is one of well-studied examples of merging clusters, located at 19$^\mathrm{h}$20$^\mathrm{m}$45$^\mathrm{s}$.3, +43$^{\circ}$57$'$43$''$ (J2000.0). This nearby ($z = 0.0557$, Struble \& Rood 1999) massive cluster is a richness 1 (Abell et al. 1989) cluster of RS-type ``cD'' (Struble \& Rood 1982), but the central cD galaxy is not very dominant, as it is classified as BM type II-III (Bautz \& Morgan 1970).  A2319 has been extensively studied in the optical ,radio and X-ray bands in the past decades. Optical analyses of the bright galaxies in A2319 (e.g., Faber \& Dressler 1977; Oegerle et al. 1995 ; Tr\`evese et al. 2000) suggested that it consists of two separate components superimposed along the line of sight: a subcluster (hereafter A2319B) around the second brightest galaxy, at about $10'$ northwest (NW) of the main concentration (hereafter A2319A).

The X-ray emission of A2319 is also asymmetric, with an elongated structure to the NW which coincides with the subcluster. The most prominent feature revealed by the Chandra observations is a sharp cold front southeast (SE) of the cD galaxy, there is also a cool arm extending around the cluster center from the tip of the cold front in the general direction of A2319B (O$'$Hara et al. 2004; Govoni et al. 2004). The peak of X-ray emission is prominently deviated from the cD galaxy, suggesting that the cluster is not dynamically relaxed. On the basis of the ASCA data, Markevitch (1996) revealed that the cluster appears largely isothermal in the central region, except for the cooler spots coincident with A2319B. Average temperature of whole cluster is 10.0$\pm$0.7 keV, consistent with the Einstein measurement (David et al. 1993), while the subcluster has a lower temperature of 8.4$\pm$1.2 keV.
%and a weak cooling flow was also reported (Edge et al. 1992).

%Observations in the hard X-ray band
%were also performed by Beppo-SAX (Molendi et al. 1999), RXTE (Gruber
%\& Rephaeli 2002), and Swift-BAT (Ajello et al. 2009).

In addition, A2319 exhibits an extended and powerful radio halo (Harris \& Miley 1978) with an irregular morphology, well correlated with the thermal X-ray distributions (Feretti et al. 1997; Govoni et al. 2001). Based on the Westerbork Synthesis Radio Telescope (WSRT) image at 20 cm, Feretti et al. (1997) showed the irregular halo structure that is very different from the prototype Coma C (Giovannini et al. 1993). Major body of the halo is elongated in the northeast (NE) to southwest (SW) direction, and there is also a NW extension, which is associated with the halo found by Harris \& Miley (1978). The relativistic non-thermal electrons responsible for the radio halo are more likely to be accelerated by the turbulence (Sugawara et al. 2009), which is probably excited by the past merger activity (Takizawa 2005).

As the main tracers of cluster dynamics, only about 140 bright galaxies  have been spectroscopically confirmed to reside in A2319. For a better understanding of dynamical structure in A2319, faint member galaxies should also be taken into account. This paper will present our multicolor photometry of the galaxies in A2319 with the Beijing-Arizona-Taiwan-Connecticut (hereafter BATC) system. The aim of the BATC cluster survey is to pick up faint member galaxies in some nearby ($z < 0.1$) clusters in different dynamic states on the basis of the spectral energy distributions (SEDs) obtained with 15 intermediate-band filters in the optical band (Yuan et al. 2001), and then investigate the dynamical substructures and the properties of member galaxies (Yuan et al. 2003; Zhang et al. 2010; Liu et al. 2011; Pan et al. 2012; Tian et al. 2012). A2319 is selected into the BATC target list as representative of powerful cluster-scale radio halo. Fig.~1 shows the smoothed contours of the ROSAT X-ray image and the radio map at 1.4 GHz from the NRAO VLA Sky Survey (NVSS), superimposed on the BATC $d$-band image.

\vspace{3mm}
\centerline{\framebox[17.5cm][c]{Fig.1: The BATC-$d$ band image, superimposed with the contours of ROSAT and NVSS brightnesses.}}
\smallskip

The rest of this paper is organized as follows: In section 2, we describe the photometric observations and data reduction. In section 3, spatial distribution, localized velocity structure and merger dynamics of A2319 are analyzed on the basis of the galaxies with known spectroscopic redshifts. A mixture-modeling algorithm, known as KMM algorithm (Ashman et al. 1994), is applied to the cluster for dividing the subcluster from the main cluster and a two-body model is also used to investigate the state of evolution. Section 4 presents the selection of faint member galaxies of A2319 by using the photometric redshift technique. In section 5, on the basis of the enlarged sample of member galaxies , we investigate the spatial distribution and star formation properties for member galaxies. Finally, we summarize our work in section 6. The $\Lambda$CDM cosmological parameters of $H_0=71$ \kms Mpc$^{-1}$, $\Omega_m=0.27$ and $\Omega_\Lambda=0.73$ are used throughout this paper.

\section{OBSERVATIONS AND DATA REDUCTION}

The BATC multicolor photometric observations were taken with the 60/90 cm f/3 Schmidt Telescope of the National Astronomical Observatories, Chinese Academy of Sciences (NAOC), located at the Xinglong site at an altitude of 900m. The photometric system contains 15 intermediate-band filters, namely, $a-k$, $m-p$, covering almost the whole optical wavelength range from $\sim$ 3000 to 10000 \AA. All these filters are especially designed to avoid bright night sky emission lines. The transmission curves of the BATC filters can be seen from Xia et al.(2002). Before October 2006, a Ford 2048 $\times$ 2048 CCD camera was equipped at the focal plane of the telescope, and the observations were carried out in only 12 bands ($d \sim p$). The field of view was $58'\times 58'$ with a spatial scale of $1.''7$ pixel$^{-1}$. To pursue a better spatial resolution and higher sensitivity in blue bands ($a-c$), a new E2V 4096 $\times$ 4096 CCD with a high quantum efficiency of 92.2\% at 4000\AA\  was equipped afterwards. This new CCD camera has a larger field of view ($92'\times 92'$) with a spatial scale of $1.''36$ pixel$^{-1}$. The pixel sizes for the old and new CCD cameras are 15$\mu m$ and 12$\mu m$, respectively.
%and the ratio is 5:4, exactly consistent with the ration of their spatial scales.
The details of the NAOC Schmidt Telescope, CCD camera and data-acquisition system can be found elsewhere (Zhou et al. 2001; Yan et al. 2000).

From 2003 January to 2008 May, images for A2319 in 15 bands has been obtained with the total exposure time more than 52 hours (see the details in Table 1). Based on the automatic data-processing software, PIPELINE I, which is designed specially for BATC multicolor photometry (Fan et al. 1996; Zhou et al. 2001), we carried out the standard procedures of bias subtraction, flat-field correction and position calibration. The cosmic rays and bad pixels were corrected by comparing multiple images.  The technique of integral pixel shifting was applied in the image combination. Since multiple exposures were obtained at various epochs, there existed the position offset and rotation to some extent in each frame.  We selected one as the reference image. For each of  other images in the same filter, relative position deviation and rotation angle were estimated, and an aligned image was produced by shifting, rotating, and interpolating operations before image combination.  
%The weighted average flux method is taken in image combination.

On the basis of the DAOPHOT kernel (Stetson 1987), the photometry package, PIPELINE II, is specially developed for detecting sources and measuring the flux within a given aperture in the combined BATC images (Zhou et al. 2003).
%As the ratio of pixel size between the old and new CCDs is 5:4, we should make the pixel size of the $a-c$ images identical with $d \sim p$ images, an aperture radius of 5 pixels (i.e. $r = 1.''36 \times 5 = 6.''8$) %is adopted for the images in three bluer bands ($a-c$), a radius of 4 pixels %( $r = 1.''7 \times 4 = 6.''8$) for the images in 12 redder bands ($d \sim %p$). However,
Due to the low Galactic latitude of the A2319 field ($\simeq13^{\circ}$), there are more than 30,000 objects detected in our viewing field, among which the stars within the Milky Way are predominated. Furthermore, this field appears more crowded due to large seeings of the combined images. For avoiding the overlapping contamination rate in aperture photometry, we elect to adopt  smaller apertures relative to our previous studies (Zhang et al. 2010, 2011): a radius of 4 pixels for $a-c$ bands, and a radius of 3 pixels for the other 12 bands ($d - p$). Flux calibrations in the $d-p$ bands are performed by using the Oke-Gunn (Gunn \& Stryker 1983) primary flux standard stars (HD19445, HD84937, BD+26d2606 and BD+17d4708), which were observed during photometric nights (Zhou et al. 2001). For obtaining the \emph{relative} SEDs of the BATC-detected sources, we take the model calibration method which has been developed on the basis of the stellar SED library especially for the BATC large-field multicolor system (Zhou et al. 1999). 

Our previous studies show that the model calibration method works well for the low-latitude fields where the interstellar extinction for these bright stars is not severe (Zhang et al. 2010; Liu et al. 2011; Pan et al. 2012; Tian et al. 2012). Despite of enhanced Galactic extinction in the A2319 field, the extinction difference between the bright stars within the BATC viewing field can be negligible . We carefully specify the initial estimates of zero-point magnitudes considering the severe Galactic extinction, and run our calibration code in iterative mode. The goodness of this model calibration can by checked by the subsequent application of photometric redshift technique to the calibrated SEDs of the galaxies with known spectroscopic redshifts.  Our calibration method may result in reasonable colors (relative magnitudes) though it might yield a systematical offset for the zero-point corrections, when compared with traditional calibration based on photometry of standard stars. 

As mentioned above, the crowed foreground stars and the bad seeings of the combined BATC images make the effect of overlapping contamination rather severe. The $u$-band image ($\lambda_{\rm eff}=3551$ \AA ) with a seeing of $1.''7$ was obtained on September 19, 2012, with the Bok 2.3m telescope at Kitt Peak, which belongs to the Steward Observatory, University of Arizona. The scale of CCD detector is $0.''455$ pixel$^{-1}$. The exposure time is 150 seconds, corresponding to a limiting magnitude of 22.$^m$5, about 1.5 mag deeper than the BATC $a$-band photometry. This $u$-band image will be taken to distinguish the overlapping contamination and exclude the fake candidates of member galaxies. The software SExtractor (version 2.5.0, Bertin \& Arnouts 1996) is taken to produce a catalogue of objects detected in the $u$-band image. A fixed aperture of $r=5.''2$ is adopted, compatible with the aperture for the BATC photometry. As a result, the position, aperture magnitude, and the star-galaxy classifier S/G (0=Galaxy, 1=Star) for all sources detected are obtained.

\def\baselinestretch{1.0}
\begin{table}[ht]
\begin{center}
%\small
%\begin{minipage}{140mm}
%\caption{ Parameters of the BATC filters and the observational statistics
%of A2319 }
%\end{minipage}
\vskip 1cm
\noindent {Table 1. Parameters of the Bok $u$-band and BATC filters and the observation log of A2319}
\vskip 1mm
\small
\begin{tabular}{ccccccccc}   \hline   \hline
\noalign{\smallskip}
 Telescope & Filter & $\lambda_{eff}$ & FWHM &
Exposure &  Number of & Seeing$^a$  &Limiting \\
   &  name & (\AA) & (\AA) &
(second) & Images     & (arcsec)  &  mag \\
\noalign{\smallskip}   \hline \noalign{\smallskip}
   Bok 2.3m  & u &  3551 & 640 & 150 &  1   & 1.71 &  22.5 \\
\noalign{\smallskip}   \hline \noalign{\smallskip}
 Schmidt 60/90cm & a  & 3369 & 222 & 26400 & 22 & 3.83 & 21.0 \\
  (BATC) & b  & 3921 & 291 & 15600 & 13 & 4.20 & 21.0 \\
  & c  & 4205 & 309 &  6300 &  7 & 4.63 & 20.5 \\
   & d  & 4550 & 332 & 25200 & 21 & 4.03 & 20.5 \\
   & e  & 4920 & 374 & 15600 & 13 & 3.83 & 20.0 \\
   & f  & 5270 & 344 & 15600 & 13 & 4.51 & 20.0 \\
  & g  & 5795 & 289 &  7200 &  6 & 4.36 & 20.0 \\
 & h  & 6075 & 308 & 11400 & 17 & 4.65 & 19.5 \\
  & i  & 6660 & 491 &  4800 &  4 & 4.25 & 19.5 \\
  & j  & 7050 & 238 &  6000 &  5 & 4.50 & 19.0 \\
  & k  & 7490 & 192 &  7200 &  6 & 4.32 & 19.0 \\
   & m  & 8020 & 255 & 10800 &  9 & 4.66 & 19.0 \\
   & n  & 8480 & 167 & 13200 & 11 & 4.48 & 19.0 \\
 & o  & 9190 & 247 &  4800 &  4 & 3.52 & 18.5 \\
  & p  & 9745 & 275 & 20400 & 17 & 4.60 & 18.5 \\
\noalign{\smallskip}
\hline
\end{tabular}
\end{center}
\vskip -3mm
\hskip 19.5mm
\parbox{120mm} {$^a$ This column lists the seeing of the combined image.}
%\tablecomments{0.96\textwidth}{$^a$ This column lists the seeing of the combined image.}
% \tablenotetext{a}{This column lists the seeing of the combined image.}
\end{table}

\def\baselinestretch{1.15}

%\vspace{4mm}
%\centerline{\framebox[9.5cm][c]{Table 1: The observation log}}
%\smallskip

\section{ANALYSES OF GALAXIES WITH KNOWN SPECTROSCOPIC REDSHIFTS}

\subsection{Sample I: Spectroscopically Confirmed Member Galaxies }

To study the properties of A2319, all the normal galaxies with known spectroscopic redshifts ($z_{\rm sp}$) in the field of $58'\times58'$ centered on A2319 are extracted from the NASA/IPAC Extragalactic Database (NED). There are 142 galaxies with the redshifts between 0.03 and 0.20, with a sharp peak at $z \sim 0.055$. Using the standard iterative 3$\sigma$-clipping algorithm (Yahil \& Vidal 1977), 138 galaxies are found to have $0.035 < z_{sp} < 0.075$. In order to refine the membership allocation, we use the ``shifting-gapper'' method (Fadda et al. 1996) to reject surviving interlopers. Left panel of Fig. 2 shows the plot of velocity versus clustercentric distance. The interlopers can be identified by applying the fixed-gap method to a bin shifting along the clustercentric distance. We adopt a gap of 1000 \kms and a bin of 0.4 Mpc (or a larger bin in order to have at least 15 galaxies). As shown in Fig. 2(a), there are 10 interlopers which are denoted by the open circles. As a result, 128 galaxies are selected as the member galaxies of A2319, which we refer to as `sample I'.  The information about the position and spectroscopic redshift for sample I is listed in Table 2. The distribution of spectroscopic redshifts for these 142 galaxies is shown in Fig.~2(b), and the histogram of  rest-frame velocities for the 128 member galaxies in sample I is also given in the embedded panel. The majority of spectroscopic redshifts were contributed by Oegerle et al. (1995). It should be pointed out that previous spectroscopy mainly focused on the central bright galaxies, against the faint galaxies in outer region. 

For quantifying the distribution of radial velocities, we apply the ROSTAT software (Beers et al. 1990) to calculate two resistant and robust estimators, namely the biweight location ($C_{\rm BI}$) and scale ($S_{\rm BI}$), which are analogous to the mean value and the standard deviation. For sample I, we derive $C_{\rm BI}=16437^{+152}_{-153}$ \kms, and $S_{\rm BI}=1713^{+97}_{-74}$ \kms. The errors of these two biweight estimators correspond to 68\% confidence interval, and they are calculated by bootstrap resamplings of 10,000 subsamples of the velocity data. Oegerle et al. (1995) obtained similar biweight location ($C_{\rm BI}=16446\pm164$ \kms) and velocity dispersion ($S_{\rm BI}=1770^{+121}_{-101}$ \kms). 

The rest-frame velocity ($V$) can be derived from spectroscopic redshifts ($z_{\rm sp}$) by $V = c(z_{\rm sp}-{z}_{c})/(1+{z}_{c})$, where $c$ is light speed, and ${z}_{c}$ is the cluster redshift with respect to cosmic background radiation. Using the ``shifting-gapper'' method, Fadda et al. (1996) obtained a sample of 118 member galaxies, and derived a velocity dispersion of $1545^{+95}_{-77}$ \kms. We take the NED-given cluster redshift, ${z}_{c} = 0.0557$, and obtain a biweight location of $C_{\rm BI}=-247^{+144}_{-145}$ \kms and a biweight scale of $S_{\rm BI}=1622^{+91}_{-70}$ \kms (i.e., intrinsic velocity dispersion). It is certain that our estimate of velocity dispersion is based on a larger sample, and thus should be more reliable. It is confirmed that A2319 has a very large velocity dispersion, almost twice the typical dispersion of a relaxed cluster, indicating a merger dynamics in A2319.

\def\baselinestretch{0.9}
\begin{deluxetable}{cccccc|cccccc}
\tabletypesize{\scriptsize}   
%\tabletypesize{\tiny}   
%\tabletypesize{\small}
%\scriptsize
%\rotate  
\tablenum{2}
\tablecaption{Catalog of 128 Spectroscopically Confirmed Member Galaxies in A2319
\label{tab2}}
\tablewidth{0pt}
%\vskip -50mm
%\tabcolsep 0.35mm  
\tabcolsep 1.8mm
\tablehead{
\colhead{No.}&\colhead{R.A.} &\colhead{Decl.} & \colhead{$z_{\rm sp}$} & \colhead{Sub.} & \colhead{Ref.}
& \colhead{No.}&\colhead{R.A.} &\colhead{Decl.} & \colhead{$z_{\rm sp}$} &\colhead{Sub.} & \colhead{Ref.}
}
%\vskip -50mm
\startdata

  1  & 290.1720833 & 43.9547222& 0.060775& B & (1)& 65 & 290.3416667 & 44.0677778& 0.054244& A & (1) \\
  2  & 290.2108333 & 43.9594444& 0.061779& B & (1)& 66 & 290.1512500 & 44.1127778& 0.061019& B & (2) \\
  3  & 290.1870833 & 43.9802778& 0.053083& A & (2)& 67 & 290.1600000 & 44.1213889& 0.062126& B & (1) \\
  4  & 290.1758333 & 43.9436111& 0.049844& A & (1)& 68 & 290.1395833 & 44.1213889& 0.064935& B & (1) \\
  5  & 290.1387500 & 43.9644444& 0.048677& A & (1)& 69 & 290.4108333 & 43.9175000& 0.048487& A & (1) \\
  6  & 290.2462500 & 43.9797222& 0.061382& B & (2)& 70 & 290.4233333 & 43.9633333& 0.049471& A & (1) \\
  7  & 290.1866667 & 44.0097222& 0.066309& B & (2)& 71 & 290.2529167 & 44.1263889& 0.056289& A & (1) \\
  8  & 290.2654167 & 43.9522222& 0.050512& A & (1)& 72 & 290.0791667 & 43.8105556& 0.064635& A & (1) \\
  9  & 290.1937500 & 44.0197222& 0.051439& A & (1)& 73 & 290.2933333 & 44.1194444& 0.055598& A & (1) \\
 10  & 290.1583333 & 44.0180556& 0.052673& A & (1)& 74 & 290.2766667 & 43.7963889& 0.046696& A & (1) \\
 11  & 290.2666667 & 43.9880556& 0.056392& A & (1)& 75 & 290.4333333 & 43.9375000& 0.057316& A & (1) \\
 12  & 290.2795833 & 43.9486111& 0.056789& A & (1)& 76 & 290.0029167 & 44.0861111& 0.058634& A & (1) \\
 13  & 290.1358333 & 44.0183333& 0.048077& A & (1)& 77 & 290.4429167 & 43.9516667& 0.045872& A & (1) \\
 14  & 290.2825000 & 43.9869444& 0.046712& A & (1)& 78 & 290.1708333 & 44.1458333& 0.057593& A & (1)  \\
 15  & 290.2883333 & 43.9691667& 0.048567& A & (1)& 79 & 289.9429167 & 44.0163889& 0.051876& A & (1)  \\
 16  & 290.2895833 & 43.9480556& 0.057153& A & (1)& 80 & 290.0800000 & 44.1425000& 0.064121& B & (1)  \\
 17  & 290.2916667 & 43.9455556& 0.054588& A & (1)& 81 & 290.0412500 & 43.7922222& 0.057416& A & (1)  \\
 18  & 290.0833333 & 43.9763889& 0.061005& B & (1)& 82 & 289.8950000 & 44.0322222& 0.048300& A & (1)  \\
 19  & 290.1525000 & 44.0380556& 0.050198& A & (2)& 83 & 289.9291667 & 44.0855556& 0.062500& B & (1)  \\
 20  & 290.1775000 & 43.8811111& 0.055165& A & (1)& 84 & 290.4116667 & 43.8038889& 0.053554& A & (1)  \\
 21  & 290.1179167 & 44.0261111& 0.065302& B & (1)& 85 & 290.2533333 & 44.1825000& 0.054137& A & (1)  \\
 22  & 290.2466667 & 43.8900000& 0.053474& A & (1)& 86 & 290.1537500 & 44.1866667& 0.044878& A & (1)  \\
 23  & 290.3045833 & 43.9594444& 0.056572& A & (1)& 87 & 290.4679167 & 43.8505556& 0.046139& A & (1)  \\
 24  & 290.1808333 & 44.0472222& 0.062393& B & (1)& 88 & 290.3337500 & 43.7516667& 0.056222& A & (2)  \\
 25  & 290.2633333 & 43.8944444& 0.054821& A & (1)& 89 & 290.2895833 & 44.1886111& 0.054955& A & (1)  \\
 26  & 290.3008333 & 43.9319444& 0.051899& A & (1)& 90 & 290.0420833 & 43.7469444& 0.052893& A & (1)  \\
 27  & 290.2954167 & 44.0016667& 0.048093& A & (1)& 91 & 290.2054167 & 44.2025000& 0.065946& B & (1)  \\
 28  & 290.2950000 & 43.9175000& 0.056055& A & (1)& 92 & 290.3220833 & 43.7327778& 0.056933& A & (1)  \\
 29  & 290.3120833 & 43.9525000& 0.047253& A & (1)& 93 & 290.5183333 & 44.0413889& 0.054698& A & (1)  \\
 30  & 290.3179167 & 43.9488889& 0.064024& B & (1)& 94 & 290.4837500 & 43.8275000& 0.058037& A & (1)  \\
 31  & 290.2808333 & 43.8952778& 0.049514& A & (1)& 95 & 290.0445833 & 44.1927778& 0.058881& A & (1)  \\
 32  & 290.2100000 & 44.0550000& 0.051869& A & (1)& 96 & 290.3245833 & 43.7272222& 0.054441& A & (2)  \\
 33  & 290.2970833 & 43.9072222& 0.056255& A & (1)& 97 & 290.3637500 & 44.1830556& 0.046345& A & (1)  \\
 34  & 290.2054167 & 44.0563889& 0.062760& B & (1)& 98 & 290.4350000 & 43.7786111& 0.060152& A & (1)  \\
 35  & 290.0562500 & 43.9655556& 0.052743& A & (1)& 99 & 290.3079167 & 43.7208333& 0.050562& A & (2)  \\
 36  & 290.3183333 & 43.9280556& 0.051529& A & (1)&100 & 290.4895833 & 43.8252778& 0.054281& A & (1)  \\
 37  & 290.1850000 & 43.8605556& 0.058160& A & (1)&101 & 290.1095833 & 44.2163889& 0.053317& A & (1)  \\
 38  & 290.3150000 & 43.9150000& 0.048990& A & (2)&102 & 289.8566667 & 44.0850000& 0.051772& A & (1)  \\
 39  & 290.2829167 & 43.8841667& 0.052139& A & (2)&103 & 290.1625000 & 43.6694444& 0.060085& A & (1)  \\
 40  & 290.3270833 & 43.9244444& 0.050231& A & (2)&104 & 290.3904167 & 44.2263889& 0.058287& A & (1)  \\
 41  & 290.3433333 & 43.9530556& 0.059478& A & (1)&105 & 290.5912500 & 43.8747222& 0.063304& A & (2) \\
 42  & 290.3233333 & 44.0194444& 0.050555& A & (1)&106 & 289.9716667 & 43.6986111& 0.046262& A & (1) \\
 43  & 290.0766667 & 44.0455556& 0.052760& A & (2)&107 & 290.3462500 & 43.6769444& 0.052003& A & (2) \\
 44  & 290.3462500 & 43.9355556& 0.062080& B & (1)&108 & 290.5350000 & 43.7747222& 0.058480& A & (1) \\
 45  & 290.3387500 & 43.9144444& 0.054488& A & (1)&109 & 290.5650000 & 43.8055556& 0.053384& A & (1) \\
 46  & 290.3437500 & 43.9119444& 0.059521& A & (1)&110 & 290.2133333 & 43.6430556& 0.055261& A & (1) \\
 47  & 290.2320833 & 44.0838889& 0.065759& B & (1)&111 & 290.1458333 & 43.6352778& 0.050038& A & (1) \\
 48  & 290.3004167 & 44.0677778& 0.055799& A & (1)&112 & 290.4737500 & 43.6994444& 0.049894& A & (1) \\
 49  & 290.0987500 & 43.8452778& 0.057523& A & (1)&113 & 290.6333333 & 44.0608333& 0.049307& A & (2) \\
 50  & 290.3433333 & 44.0380556& 0.064721& B & (1)&114 & 290.6529167 & 43.9211111& 0.045982& A & (1) \\
 51  & 290.1379167 & 44.0919444& 0.064048& B & (2)&115 & 290.5995833 & 43.8000000& 0.050425& A & (1) \\
 52  & 290.3579167 & 44.0263889& 0.060739& B & (2)&116 & 289.8570833 & 43.7127778& 0.058697& A & (1) \\
 53  & 290.3783333 & 43.9350000& 0.050595& A & (2)&117 & 290.3504167 & 43.6347222& 0.046812& A & (1) \\
 54  & 290.3566667 & 43.8880556& 0.055091& A & (2)&118 & 290.1162500 & 43.6055556& 0.051425& A & (1) \\
 55  & 290.3825000 & 43.9955556& 0.045778& A & (3)&119 & 290.6087500 & 44.1611111& 0.059618& A & (1) \\
 56  & 290.1720833 & 44.1052778& 0.061576& B & (1)&120 & 290.1620833 & 43.5947222& 0.052563& A & (1) \\
 57  & 290.3633333 & 44.0333333& 0.052600& A & (1)&121 & 290.6929167 & 44.0958333& 0.047860& A & (1)  \\
 58  & 290.3120833 & 44.0805556& 0.061869& B & (1)&122 & 290.5529167 & 44.2536111& 0.057023& A & (4) \\
 59  & 290.2204167 & 43.8147222& 0.054097& A & (1)&123 & 290.3670833 & 43.5888889& 0.055542& A & (2) \\
 60  & 290.3662500 & 43.8836111& 0.044514& A & (1)&124 & 290.1429167 & 43.5641667& 0.059424& A & (5) \\
 61  & 290.0412500 & 43.8561111& 0.054391& A & (1)&125 & 290.6795833 & 43.7594444& 0.058907& A & (1) \\
 62  & 290.2979167 & 43.8341667& 0.060258& A & (2)&126 & 290.6783333 & 43.6925000& 0.053317& A & (1) \\
 63  & 290.3579167 & 44.0505556& 0.053934& A & (1)&127 & 290.6208333 & 43.6444444& 0.057206& A & (1) \\
 64  & 290.3491667 & 43.8644444& 0.047403& A & (1)&128 & 290.6879167 & 44.2383333& 0.049461& A & (6)

%\noalign{\smallskip}
%\noalign{\smallskip}
\enddata

%\tablecomments{0.86\textwidth} {$^{a}$ Effective wavelengths of the filters.}
% \tablenotetext{a}{This column lists the seeing of the combined image.}

%\noindent \textbf{References}. (1) Oegerle et al. 1995; (2) Faber \& Dressler 1977; (3) De Vaucouleurs et al. 1991;
%(4) Marzke et al. 1996; (5) Nakanishi et al. 1997; (6) Huchra et al. 2012.

%\vskip -2mm
%\hskip 6mm \parbox{50mm} {$^a$ seeing of the combined image}

\vskip 1mm
\parbox{155mm} {\textbf{Ref}. (1) Oegerle et al. 1995; (2) Faber \& Dressler 1977; (3) De Vaucouleurs et al. 1991;
(4) Marzke et al. 1996; (5) Nakanishi et al. 1997; (6) Huchra et al. 2012.}

\end{deluxetable}

\def\baselinestretch{1.15}

%\vspace{4mm}
%\centerline{\framebox[14.5cm][c]{Table 2: Catalog of 138 spectroscopically confirmed member galaxies in A2319}}
%\smallskip

%\begin{tabular}{ll}
%Column 1: & Number of the galaxies (sorted by the distance to cluster
%center);\\
%Column 2: & R.A. in 2000 epoch, in degrees, given by NED;\\
%Column 3: & Declination in 2000 epoch, in degrees, given by NED; \\
%Column 4: & Spectroscopic redshift in NED;\\
%Column 5: & The literature giving the Spectroscopic redshift.\\
%%Column 6 - 18: & Photometric magnitudes in 13 BATC filter bands.
%%The value of 0.0 \\
%%& means out of field;\\
%%Column 19 - 23: & Aperture-corrected photometric magnitudes in five SDSS
%filter bands.
%\end{tabular}

\vspace{4mm}
\centerline{\framebox[16.5cm][c]{Fig.~2: Distributions of the spectroscopic redshifts and radial velocities for 138 member galaxies}}
\smallskip

\subsection{ Spatial Distribution and Localized Velocity Structure}

Faber \& Dressler (1977) pointed out a gap in the velocity distribution (see Fig.~1 therein) of A2319, and suggested that A2319 is really two clusters (A2319A and A2319B) viewed in projection. With four times larger sample of member galaxies, the velocity distribution of sample I does not show clear evidence for two merging subclusters (see Fig.~2). The left panel of Fig.~3 presents the spatial distribution of sample I with respect to the NED-given central position of A2319 ($19^h21^m08^s.8$, $+43^{\circ}57'30''$; J2000.0). We superpose the contour map of surface density, which has been smoothed by a Gaussian window with $\sigma = 1.'6$. The brightest cluster galaxy (BCG), namely, CGCG~230-007, is denoted by the filled triangle. Spatial distribution of sample I remarkably deviates from spherical symmetry. Irregular contour map shows that the main body is elongated in the SE-NW direction, which seems to be consistent with the X-ray brightness distribution and the radio halo morphology (Feretti 1997; Harris \& Miley 1978). Fig.~3(a) reveals six satellite-like clumps (i.e., possible substructures, namely B, C, D, E, and F) surrounding the main concentration A.

%These clumps may correspond to locations of the possible cluster radio galaxies (Feretti 1997).

However, apparent clumps may be the enhancements simply due to the projection effect. To verify the presence of substructures, we employ a quantitative statistics method called the ``$\kappa$-test" (Colless \& Dunn 1996). In this test, a statistic, $\kappa_n$, is defined to quantify the local deviation on the group scale $n$. A larger $\kappa_n$ implies a greater probability that local velocity distribution for $n$ neighboring galaxies  differs from overall velocity distribution. The probability, $P(\kappa_n > \kappa^{\rm obs}_n)$, can be estimated by Monte Carlo simulations with randomly shuffling velocities. Table 3 gives the results of $P(\kappa_n > \kappa^{\rm obs}_n)$ in $\kappa$-test with different neighbor sizes.  For all cases, 1000 simulations  are performed for probability estimate.

%\input{tab3.tex}
%table 3
\begin{table}[ht]
%\caption[]
\begin{center}
%\small
\noindent {Table 3. Result of $\kappa$-test for 128 member galaxies in sample I}
%\begin{minipage}{125mm}
%\caption{\small  Result of $\kappa$-Test for 128 Member Galaxies in sample I}
%\end{minipage}
\vskip 2mm
\begin{tabular}{cccccccc}   \hline   \hline
\noalign{\smallskip}
%Sample size
 Neighbor size ($n$) &      6 &    7&      8&      9&      10&     11&     12\\
\noalign{\smallskip} \hline \noalign{\smallskip}
$P(\kappa_n>\kappa_n^{\rm obs})$  &
  6.0\% &  3.2\%&  3.8\%&  3.3\%&  3.0\%&  1.2\%&  1.3\%\\
%Sample II & 5.6\%  &  2.7\% &  6.1\%&  8.6\%&  5.3\%&  5.3\%&  7.6\%\\
\noalign{\smallskip}   \hline
\end{tabular}
\end{center}
\end{table}

%\vspace{4mm}
%\centerline{\framebox[14cm][c]{Table 3:  The probabilities $P(\kappa_n > \kappa^{\rm obs}_n)$ in the $\kappa$-tests for samples I and II}}
%\smallskip

%\input{tab3.tex}

The existence of substructure in A2319 is strongly supported by the variation of local velocity  in a wide range of neighbor sizes $n$. As $7 \leq n \leq 12$, the probability $P(\kappa_n > \kappa^{\rm obs}_n)$ are less than 5\%. 
%In particular, the probability $P(\kappa_n > \kappa^{\rm obs}_n)$ is nearly zero when $n=5$, %which means $n=5$ is the optimum scale on which the substructure appears most obvious. 
The bubble plot is given in the right panel of Fig.~3, showing the location of localized variation using $n=7$ nearest neighbors. The bubbles centered on member galaxies have their sizes that are proportional to $-log[P(D>D_{\rm obs})]$. Therefore the clustering of large bubbles is a good tracer of dynamically substructure. A close comparison of two panels shows that the bunch of large bubbles, located at $\sim 10'$ NW to the main concentration A2319A, is associated with clump B. Our $\kappa$-test confirms the presence of subcluster A2319B (Faber \& Dressler 1977; Oegerle et al. 1995). 
%Additionally, a clustering of 5 large bubbles occurs at the position of clump C, which indicates a probable dynamical substructure. Because of the limited number of galaxies in clump C,
%it is rather arbitrary to claim this substructure on the basis of the spectroscopic sample only. 
%We have yet to verify this possible substructure with a larger sample of member galaxies. 
Since no sign of bubble clustering can be seen in the positions of clumps C, D, E, and F,  these clumps should be unreal substructures that are produced by projection effect.

\vspace{4mm}
\centerline{\framebox[14.0cm][c]{Fig. 3: Spatial distribution and bubble plot for sample I}}
\smallskip

\subsection{ Dynamical Properties of A2319 }

In order to study the gravitational interaction between the main cluster (A2319A) and the subcluster (A2319B), the 128 galaxies in sample I ought to be partitioned.  It is relatively easy to partition the galaxies close to the group centers. However, for the galaxies located halfway between the group centers, the partition might become a rather ambiguous task. For achieving a robust partition, we apply a technique of mixture modeling, namely the KMM algorithm, a maximum-likelihood algorithm which assigns objects into groups and assesses the improvement in fitting a multi-group model over a single-group model (Ashman et al. 1994). Once the initial values of central positions and velocity dispersions for A2319A and A2319B are given, the KMM algorithm can start iterating toward the maximum-likelihood solution. After several iterations,  a convergent solution is obtained, and rates of correct allocation are 0.975 and 0.955 for A2319A and A2319B, respectively.   A2319A includes 106 galaxies, and has a biweight location of $C_{\rm BI}=15,958_{-134}^{+137}$ \kms and a biweight scale of $S_{\rm BI}=1,352_{-75}^{+86} $ \kms. A2319B contains 22 members, with $C_{\rm BI}= 18,883_{-195}^{+133}$ \kms and $S_{\rm BI}=573_{-54}^{+73} $ \kms. The significant difference in radial velocity distribution is well shown with the stripe density plot in Fig.~4(b). The $C_{\rm BI}$ difference is about 3,000 \kms, three times as large as typical velocity dispersion for a relaxed cluster, which supports the merging picture along the line of sight. The rest-frame velocity dispersions, $S_{\rm BI}/(1+z_c)$, are found to be $1,281_{-71}^{+81}$ and  $543_{-51}^{+69}$ \kms  for A2319A and A2319B, respectively.

\vspace{4mm}
\centerline{\framebox[17cm][c]{Fig.~4:  Spatial distribution and tripe density plot of radial velocities for subclusters A, and B.}}
\smallskip

Based on only 31 galaxy spectra, Faber \& Dressler (1977) derived $\bar{v}_{\rm A}= 15,882$ km~s$^{-1}$, $\sigma_{\rm A} = 873^{+131}_{-148}$ \kms, and $\bar{v}_{\rm B}= 19,074$ \kms, $\sigma_{\rm B} = 573^{+120}_{-149}$ \kms. Oegerle et al. (1995) partitioned 100 and 28 galaxies to A2319A and A2319B, respectively, and fitted with two Gaussian velocity distributions:  $\bar{v}_{\rm A}= 15,727$ \kms, $\sigma_{\rm A} = 1,324$ \kms, and $\bar{v}_{\rm B}= 18,636$ \kms, $\sigma_{\rm B} = 742$ \kms. Tr\`evese et al. (2000) reassigned these galaxies through computing the probability distributions of galaxies, and obtained $N_{\rm A}=96$, $N_{\rm B}=24$ ; $\bar{v}_{\rm A}= 15,891$ \kms, $\sigma_{\rm A} = 1,235\pm90$ \kms; $\bar{v}_{\rm B} = 18,859$ \kms, $\sigma_{\rm B} = 655\pm97$ \kms. These results were based on different decomposition methods and sample sizes. Our statistics results are based on the KMM algorithm and the updated redshift database, and is thus more reliable.

To judge whether A2319B is gravitationally bound to the main concentration A2319A, we estimate the masses of two parts by applying virial theorem. Assuming that each component is bound and that the galaxy orbits are random, the virial mass ($M_{\rm vt}$) can be estimated from the following standard formula (Geller \& Peebles 1973; Oegerle \& Hill 1994), which assumes spherical symmetry:
\begin{equation}
M_{\rm vt}=\frac{3\pi}{G}\sigma_{r}^{2}DN_{p}\left( \sum\limits_{i>j}^{N}\frac{1}{\theta_{ij}}\right)^{-1},
\end{equation}
where $\sigma_{r}$ is the line-of-sight velocity dispersion, $D$ is the cosmological distance of the cluster, $N_{p}=N(N-1)/2$ is the number of galaxy pairs, and $\theta_{ij}$ is the angular separation between galaxies $i$ and $j$. We find the virial masses of $2.32 (\pm 0.27) \times10^{15}M_{\odot}$ and $2.36 (\pm 0.51) \times10^{14}M_{\odot}$ for subclusters A2319A and A2319B, respectively. The uncertainties in virial mass estimate are computed by propagating the uncertainty on velocity dispersion and mean angular separation. The limits on bound orbits for these two subclusters can be determined by using the Newtonian criterion for gravitational binding:
\begin{equation}
V_{r}^{2}R_{p}\leq2GM_{\rm tot}\sin^{2}\alpha\cos\alpha,
\end{equation}
where $V_{r}$ is the relative velocity along the line of sight, $R_{p}$ is the projected separation, $M_{\rm tot}$ is the total mass, and $\alpha$ is the angle between the plane of sky and the line joining the two subclusters (Beers et al. 1982). We calculate the projected distance, $R_p$, using the centroids produced by the KMM partition algorithm.  The angular separation is about 9.12 arcmin, which corresponds to a projected distance of 0.606 Mpc. The resulting constraint on $V_{r}$ for bound orbits is shown in Fig.~5, as a function of the projection angle $\alpha$. The dashed curves in this plot represent the uncertainty of the Newtonian binding criterion when the uncertainty of total virial mass is taken into account.  The  vertical lines represents the relative radial velocity $V_{r} = (C_{\rm BI,B}-C_{\rm BI,A})/(1+z_c) = 2771 \pm 202 $ \kms in the rest frame of A2319.  

Following Yang et al. (2004), the probability of gravitational binding can be computed by the formula
\begin{equation}
P=\frac{1}{A}\int_0^{+\infty} p(V_r) p(\alpha |V_r) dV_r,
\end{equation}
where  $A$ is renormalization factor;  $p(V_r)$ is the probability distribution of the relative velocity which can be assumed to be a Gaussian; $p(\alpha|V_r)$ is the probability of the valid   $\alpha$ for gravitational binding at a given $V_r$, which can be derived from equation (2). Integration is over all the appropriate values of velocity. As a result, these two subclusters are gravitationally bound with a probability of $0.53 _{-0.03}^{+0.02}$. The probability uncertainty is computed when the uncertainty of total virial mass is considered. O$'$Hara et al. (2004) unveiled a  cold front with the Chandra imaging observation, and proposed a two-body merger where
A2319B merged from the southeast traveling northwest.  The finding of cold front seems to rule out a merger along the line of sight. However, our measurements of radial velocity difference
between the subclusters is very large, indicating that the merger is not taking place in the plane
of the sky. Considering the X-ray cold front, we would like to propose a merger model where the trajectory lies approximately $50\pm20^{\circ}$ out of the plane of the sky. Therefore, it is likely that these two subclusters are gravitationally bound, as shown in Fig.~5. 

%Fig.~5, A2319B is bound to A2319A if the projection angle is in the range of  $21.8^{\circ} < \alpha < 82.5^{\circ}$.

\vspace{4mm}
\centerline{\framebox[14cm][c]{Fig.~5:   Constraint on bound orbits in the ($V_{r}-\alpha$) plane for A2319A and A2319B}}
\smallskip

\section{ SELECTION OF FAINT MEMBER GALAXIES}

\subsection{ Star-Galaxy Separation}

%There is no doubt that optical spectroscopy is a reliable means for membership determination. It is still a time-consuming task to select  faint member galaxies via spectroscopy.

The technique of photometric redshift has been developed to estimate the redshifts of faint galaxies based on their SEDs with wide coverage of wavelength (Pell\'o et al. 1999; Brunner \& Lubin 2000).  To detect the faint member galaxies from the remaining BATC-detected sources, the star-galaxy separation is firstly performed. The color-color diagram is a powerful tool for separating stars and galaxies. As shown in Yuan et al. (2001), since the spectra of redshifted galaxies differ significantly from those of stars, different classes of objects occupy different regions in the color-color diagrams. Fig.~6 presents three color-color diagrams used for star-galaxy separation. The diagrams include following categories of sources: (i) all types of stars in our SED template library (denoted by triangles), (ii) morphologically various galaxies with template SEDs (denoted by open circles), (iii) the spectroscopically confirmed member galaxies in sample I (denoted by crosses), and (iv) all the remaining sources detected by the BATC photometry (denoted by dots). Five filters $b$, $f$, $h$, $o$ and $p$ are used in the diagrams. As seen from Fig.~6, the stars in the template library lie in a well-defined region stretching from top-left to bottom-right, while the spectroscopically confirmed member galaxies locate just above two dashed lines.

\vspace{4mm}
\centerline{\framebox[12cm][c]{Fig.~6:  The color-color diagrams for star-galaxy separation}}
\smallskip

The $u$-band image obtained by the Bok 2.3m telescope can be taken to distinguish the galaxies from the foreground stars. With the photometric list of the $u$-band image, produced by the software SExtractor, we cross-identify the BATC sources that satisfy the color-color selection criteria. As a result, 927 galaxies brighter than $h_{\rm BATC}=19.0$ are found to have the star-galaxy classifier S/G$<0.9$, and these extended sources are picked out for further $z_{\rm ph}$ estimate.

\subsection{Cluster Membership Selection}

The technique of photometric redshift has been extensively applied to detect faint and distant (i.e. high-$z$ ) galaxies (Bolzonella et al. 2000; Ilbert et al. 2009) and to select cluster member galaxies (Brunner \& Lubin  2000; Finoguenov et al. 2007). For a given object, with the standard SED-fitting code called HYPERZ (Bolzonella et al. 2000), the procedures for $z_{\rm ph}$ estimate have been developed especially for the BATC multicolor photometric system (Xia et al. 2002), the photometric redshift corresponds to the best fit (in the $\chi^2$-sense) between the observed SED and the template SED which has been generated by convolving the galaxy spectra with the transmission curves of the BATC filters (Yuan et al. 2003; Yang et al. 2004; Liu et al. 2011). During the SED fitting, we only take the reference templates of normal galaxies into account. For the galaxies in A2319 field at low Galactic latitudes, an excess of extinction correction should be taken into account when we apply the photometric redshift technique.  An extinction threshold of $A_V=0.2$ is initially set to cover the extinction excess and internal reddening with the same law of Milky Way (Allen 1976).  As a result, $A_{V}$ is flexible in a range from 0.2 to 0.6, with a step of 0.01. 

The 142 bright galaxies with known spectroscopic redshifts ($z_{\rm sp}$) in our field of view are unambiguously cross-identified with the BATC-detected sources.  The 15-band SEDs of these galaxies can be used to check the reliability of $z_{\rm ph}$ estimate. Fig.~7(a) shows the $z_{\rm ph}$ - $z_{\rm sp}$ comparison for these 142 galaxies. The solid line indicates $z_{\rm ph}=z_{\rm sp}$, the error bars in the $z_{\rm ph}$ determination at 68\% confidence level are also given, and the dashed lines show 2$\sigma$ redshift deviation. It is clear that our $z_{\rm ph}$ estimates are basically consistent with their $z_{\rm sp}$ values. For the galaxies with  redshifts smaller than 0.07, the average standard deviation of the redshift difference is 0.005. There exists a slight systematic offset in the $z_{\rm ph}$ domain with respective to the $z_{\rm sp}$ distribution. For the 128 member galaxies in sample I, the mean value and the standard deviation of their $z_{\rm ph}$ values are 0.0544 and 0.0045, respectively. Applying the iterative 3$\sigma$-clipping algorithm, over 95 percent of member galaxies are found to have their $z_{\rm ph}$ values within the $\pm3 \sigma$ deviation, in the range from 0.0408 to 0.0679. This $z_{\rm ph}$ range can then be taken as the selection criterion for faint member galaxies, which is denoted by the dashed lines in Fig.~7(b).

\vspace{4mm}
\centerline{\framebox[16.5cm][c]{Fig.~7: The $z_{\rm sp}$-$z_{\rm ph}$ comparison and the $z_{\rm ph}$ distribution for 142 galaxies with known redshifts}}
\smallskip

Then, we search the $z_{\rm ph}$ values for the 927 galaxies brighter than $h_{\rm BATC}=19.0$, selected by the color-color diagrams and the SExtractor classifier S/G$<0.9$, in the redshift range from 0.0 to 0.3 with a step of 0.001.  Fig.~8 shows the $z_{\rm ph}$ distribution for these galaxies, and the dashed lines correspond to the $z_{\rm ph}$ range of membership selection criterion. For these faint galaxies without spectroscopic redshifts, we calculate the average and standard deviation of the $z_{ph}$ uncertainties for four magnitude bins ($15.5<i<16.5$, $16.5<i<17.5$, $17.5<i<18.5$,and $18.5<i<19.0$), their standard errors are 0.0012, 0.0018, 0.0035, and 0.0106, respectively, quite similar to that for A2589 (see Fig.~7 in Liu et al. 2012).  The degree of SED fitting tends to be worse for the faint galaxies with greater magnitude errors.

Due to severe effect of overlapping contamination in the BATC images, the light within photometric aperture might have been contributed by foreground stars or neighboring galaxies.  It is necessary to take a cautious eyeballing check one by one  to exclude the false candidates whose BATC SEDs are contaminated to some degree, by comparing the details in the Bok $u$-band image and the BATC images. As a result, 233 faint member candidates (including 216 early- and 17 late-type galaxies) are obtained. Table 4 presents their information about the BATC-detected celestial coordinates, photometric redshifts, and  morphological classifications of the best-fit templates. Thus, after combining with the 128 member galaxies in sample I (including 118 early- and 10 late-type galaxies), we finally obtain an enlarged sample of 361 member galaxies in A2319, which is referred to as sample II.

\vspace{4mm}
\centerline{\framebox[14.5cm][c]{Fig.~8: The $z_{\rm ph}$ distribution of the 927 galaxies brighter than $h_{\rm BTAC}=19.0$}}
\smallskip

%\vspace{4mm}
%\centerline{\framebox[12cm][c]{Table 4: SED Catalog of 233 newly-selected member galaxies}}
%\smallskip

 \def\baselinestretch{0.98}
 \begin{deluxetable}{ccccc|ccccc}
%\tabletypesize{\footnotesize} %% 表格字体大小
\tabletypesize{\scriptsize}
%\tabletypesize{\tiny}
%\tabletypesize{\small}
%\centering
%\rotate     %% 把表格转过90度。
\tablenum{4}
\tablecaption{Catalog of 233 Newly-Selected Member Galaxies in
A2319}
\label{tab4}
\tablewidth{0pt}
\tabcolsep 2.1mm
\tablehead{
\colhead{No.}&\colhead{R.A.} &\colhead{Decl.} & \colhead{$z_{\rm ph}$} & \colhead{$T$} &
\colhead{No.}&\colhead{R.A.} &\colhead{Decl.} & \colhead{$z_{\rm ph}$} & \colhead{$T$}
}
\startdata

  1 & 290.5707917&  43.5564167&  0.050&  1&   61 & 290.1465417&  43.7897222&  0.043&  2  \\
  2 & 290.3098750&  43.5772222&  0.061&  1&   62 & 289.6606667&  43.7927778&  0.059&  1  \\
  3 & 290.1505417&  43.7240833&  0.045&  5&   63 & 290.2554583&  43.7894722&  0.053&  2  \\
  4 & 290.4315833&  43.9894167&  0.063&  2&   64 & 289.6908750&  43.7990556&  0.062&  1  \\
  5 & 290.2465417&  44.0871389&  0.054&  1&   65 & 290.2033333&  43.8037778&  0.053&  1  \\
  6 & 289.8410417&  44.1006389&  0.056&  2&   66 & 290.5016250&  43.8049722&  0.055&  1  \\
  7 & 289.7957500&  44.1354722&  0.051&  1&   67 & 290.2696250&  43.8133056&  0.054&  1  \\
  8 & 290.2976250&  44.1460278&  0.057&  1&   68 & 289.9811667&  43.8158889&  0.055&  1  \\
  9 & 290.4627083&  44.3364722&  0.050&  1&   69 & 290.1829583&  43.8182500&  0.054&  1  \\
 10 & 290.0179167&  43.5144444&  0.056&  1&   70 & 290.7453750&  43.8215000&  0.045&  1  \\
 11 & 290.1404167&  43.5164167&  0.062&  1&   71 & 289.7741667&  43.8376944&  0.041&  2  \\
 12 & 290.2332917&  43.5324444&  0.054&  1&   72 & 290.2718750&  43.8383889&  0.055&  2  \\
 13 & 290.5399583&  43.5397778&  0.054&  1&   73 & 290.2440833&  43.8392778&  0.049&  1  \\
 14 & 289.8405833&  43.5606667&  0.041&  1&   74 & 290.5167500&  43.8441667&  0.055&  1  \\
 15 & 290.0956667&  43.5679722&  0.054&  1&   75 & 290.1050833&  43.8566944&  0.052&  1  \\
 16 & 290.3252083&  43.5785278&  0.061&  1&   76 & 290.1954583&  43.8585000&  0.057&  1  \\
 17 & 290.2490000&  43.5803611&  0.054&  1&   77 & 290.2132500&  43.8629444&  0.054&  2  \\
 18 & 289.9237917&  43.5839167&  0.054&  1&   78 & 290.2357500&  43.8658889&  0.055&  1  \\
 19 & 290.2033750&  43.5867222&  0.048&  7&   79 & 290.2887083&  43.8698056&  0.059&  1  \\
 20 & 290.5572917&  43.5882778&  0.052&  1&   80 & 289.8280000&  43.8754444&  0.054&  1  \\
 21 & 289.6702083&  43.5956667&  0.057&  1&   81 & 289.8515000&  43.8758889&  0.053&  1  \\
 22 & 290.0621667&  43.5950833&  0.056&  1&   82 & 290.3647917&  43.8849167&  0.048&  1  \\
 23 & 290.1777083&  43.5966389&  0.058&  1&   83 & 290.3200833&  43.8880556&  0.058&  1  \\
 24 & 290.2596667&  43.5970556&  0.050&  1&   84 & 290.2375000&  43.8922778&  0.054&  1  \\
 25 & 290.3559583&  43.5963611&  0.059&  1&   85 & 289.6224167&  43.8962778&  0.048&  6  \\
 26 & 290.0515000&  43.6158056&  0.048&  1&   86 & 290.6678333&  43.8885278&  0.042&  2  \\
 27 & 290.4890833&  43.6129167&  0.053&  1&   87 & 290.3679583&  43.8935833&  0.045&  1  \\
 28 & 290.5767917&  43.6131667&  0.054&  1&   88 & 290.2880417&  43.8986667&  0.055&  1  \\
 29 & 290.1737500&  43.6188611&  0.053&  1&   89 & 290.7234583&  43.8961667&  0.055&  1  \\
 30 & 289.9596250&  43.6259444&  0.053&  1&   90 & 290.3599167&  43.9017222&  0.055&  1  \\
 31 & 290.5954583&  43.6277222&  0.058&  1&   91 & 290.6178750&  43.8994444&  0.046&  1  \\
 32 & 290.3765833&  43.6486944&  0.050&  3&   92 & 290.6055833&  43.9101944&  0.057&  1  \\
 33 & 290.2935833&  43.6571667&  0.046&  1&   93 & 290.2883333&  43.9153333&  0.056&  1  \\
 34 & 290.5813750&  43.6565278&  0.051&  1&   94 & 289.9920417&  43.9206389&  0.046&  1  \\
 35 & 290.3021250&  43.6612500&  0.059&  1&   95 & 290.2546250&  43.9233333&  0.053&  1  \\
 36 & 289.8011667&  43.6670278&  0.048&  1&   96 & 290.4669167&  43.9272500&  0.050&  1  \\
 37 & 289.7814167&  43.6727222&  0.053&  1&   97 & 290.2599167&  43.9299444&  0.041&  1  \\
 38 & 290.0677500&  43.6721944&  0.057&  2&   98 & 290.7916667&  43.9339722&  0.053&  1  \\
 39 & 290.3675417&  43.6788056&  0.051&  1&   99 & 289.7853333&  43.9475278&  0.051&  2  \\
 40 & 289.8690417&  43.6881111&  0.055&  1&  100 & 290.1115417&  43.9474722&  0.057&  1  \\
 41 & 290.1201250&  43.6945000&  0.063&  1&  101 & 290.1100833&  43.9512778&  0.053&  1  \\
 42 & 290.2097917&  43.7105278&  0.052&  1&  102 & 290.2388333&  43.9515833&  0.053&  1  \\
 43 & 290.2817500&  43.7115833&  0.052&  2&  103 & 290.2942500&  43.9538611&  0.055&  1  \\
 44 & 290.4338333&  43.7219722&  0.059&  1&  104 & 290.8193750&  43.9540556&  0.050&  1  \\
 45 & 290.1207917&  43.7312222&  0.055&  1&  105 & 290.3029167&  43.9615278&  0.059&  1  \\
 46 & 290.2051250&  43.7382222&  0.056&  1&  106 & 290.3253750&  43.9653056&  0.042&  3  \\
 47 & 290.0104167&  43.7416389&  0.052&  1&  107 & 290.3345000&  43.9725833&  0.053&  1  \\
 48 & 289.6098333&  43.7450000&  0.046&  1&  108 & 290.2770417&  43.9742778&  0.055&  1  \\
 49 & 290.3342917&  43.7503056&  0.053&  1&  109 & 290.3753333&  43.9749444&  0.054&  2  \\
 50 & 289.5722083&  43.7577778&  0.057&  1&  110 & 290.3093750&  43.9775556&  0.057&  1  \\
 51 & 290.0219167&  43.7555278&  0.054&  1&  111 & 290.2000833&  43.9802222&  0.059&  1  \\
 52 & 289.8010417&  43.7667500&  0.046&  1&  112 & 290.3965000&  44.0018056&  0.057&  1  \\
 53 & 290.2874583&  43.7647778&  0.054&  1&  113 & 290.1347500&  44.0081944&  0.054&  1  \\
 54 & 290.1413750&  43.7769722&  0.042&  1&  114 & 289.6401667&  44.0114722&  0.055&  1  \\
 55 & 289.6819583&  43.7816389&  0.062&  1&  115 & 290.3554583&  44.0069167&  0.048&  2  \\
 56 & 289.6033750&  43.7825000&  0.057&  1&  116 & 290.1608333&  44.0151389&  0.059&  1  \\
 57 & 290.0545000&  43.7866389&  0.046&  1&  117 & 289.6872500&  44.0205278&  0.052&  2  \\
 58 & 290.2214167&  43.7854444&  0.055&  1&  118 & 290.7510833&  44.0185556&  0.059&  6  \\
 59 & 290.4391250&  43.7834722&  0.054&  1&  119 & 289.7245417&  44.0331944&  0.060&  3  \\
 60 & 290.4613750&  43.7857500&  0.053&  1&  120 & 289.7214167&  44.0356944&  0.049&  1

\enddata
\vskip 1mm
\parbox{125mm}{Hubble type of the best-fit template $T$: 1=E, 2=S0, 3=Sa, 4=Sb, 5=Sc, 6=Sd, 7=Irr}
%\tablecomments{155mm}{Hubble type of the best-fit template $T$: 1=E, 2=S0, 3=Sa, 4=Sb, 5=Sc, 6=Irr}
\end{deluxetable}

%\clearpage
\begin{deluxetable}{ccccc|ccccc}
%\tabletypesize{\footnotesize} %% 表格字体大小
\tabletypesize{\scriptsize}
%\tabletypesize{\tiny}
%\tabletypesize{\small}
%\centering
%\rotate     %% 把表格转过90度。
\tablewidth{0pt}
\tablenum{4}
\tablecaption{ -- continued}
%\label{tab4}}
%\tabcolsep 0.9mm
\tabcolsep 2.1mm
\tablehead{
\colhead{No.}&\colhead{R.A.} &\colhead{Decl.} & \colhead{$z_{\rm ph}$} & \colhead{$T$} &
\colhead{No.}&\colhead{R.A.} &\colhead{Decl.} & \colhead{$z_{\rm ph}$} & \colhead{$T$}
}
\startdata
121 & 290.8143750&  44.0300000&  0.053&  1&  181 & 289.9196250&  44.2634167&  0.048&  1  \\
122 & 290.6606250&  44.0356944&  0.046&  1&  182 & 290.3724583&  44.2601944&  0.043&  4  \\
123 & 290.7657500&  44.0461944&  0.053&  1&  183 & 290.1582500&  44.2645833&  0.052&  1  \\
124 & 289.5807500&  44.0596389&  0.061&  1&  184 & 290.5285000&  44.2715000&  0.050&  1  \\
125 & 289.8155833&  44.0593056&  0.048&  3&  185 & 290.5580417&  44.2845278&  0.051&  1  \\
126 & 290.6306667&  44.0546667&  0.057&  1&  186 & 290.3392917&  44.2988333&  0.052&  1  \\
127 & 290.3037500&  44.0589722&  0.047&  1&  187 & 289.9407500&  44.3052778&  0.053&  1  \\
128 & 289.9426667&  44.0629167&  0.049&  2&  188 & 290.2791667&  44.3151111&  0.053&  1  \\
129 & 290.2127083&  44.0617222&  0.063&  1&  189 & 290.4480833&  44.3146389&  0.052&  1  \\
130 & 290.4850833&  44.0905000&  0.048&  1&  190 & 289.6488333&  44.3226111&  0.056&  3  \\
131 & 290.2482500&  44.0962222&  0.046&  1&  191 & 290.6006250&  44.3179722&  0.048&  1  \\
132 & 289.6668333&  44.1062778&  0.046&  1&  192 & 290.2172083&  44.3251944&  0.045&  1  \\
133 & 290.1667083&  44.1140556&  0.055&  1&  193 & 289.6846667&  44.3311389&  0.053&  1  \\
134 & 289.9489583&  44.1203056&  0.045&  2&  194 & 290.0548333&  44.3311944&  0.052&  3  \\
135 & 290.0893333&  44.1283056&  0.054&  1&  195 & 290.6830000&  44.3290000&  0.048&  1  \\
136 & 290.7906250&  44.1305556&  0.049&  1&  196 & 290.5552083&  44.3308333&  0.054&  1  \\
137 & 289.9152917&  44.1383056&  0.048&  2&  197 & 290.0032083&  44.3379444&  0.056&  1  \\
138 & 289.6950417&  44.1431389&  0.055&  2&  198 & 290.6210417&  44.3400556&  0.053&  1  \\
139 & 289.9263750&  44.1423611&  0.047&  1&  199 & 289.9171667&  44.3496944&  0.066&  1  \\
140 & 289.8295417&  44.1455833&  0.057&  1&  200 & 289.8088333&  44.3570278&  0.047&  1  \\
141 & 289.9111667&  44.1553333&  0.053&  2&  201 & 290.6928750&  44.3512778&  0.053&  1  \\
142 & 290.6840000&  44.1554722&  0.053&  1&  202 & 290.3861250&  44.3711111&  0.056&  1  \\
143 & 290.0590417&  44.1629722&  0.053&  1&  203 & 290.0052917&  44.3881111&  0.057&  1  \\
144 & 290.1371667&  44.1629167&  0.052&  2&  204 & 290.4163333&  44.3870000&  0.056&  1  \\
145 & 290.5493333&  44.1594444&  0.059&  1&  205 & 290.3081250&  44.4064167&  0.044&  1  \\
146 & 290.2621250&  44.1632222&  0.063&  1&  206 & 289.6782083&  44.4114722&  0.057&  1  \\
147 & 290.4014583&  44.1730000&  0.048&  1&  207 & 289.9890417&  44.4426667&  0.057&  2  \\
148 & 290.7360833&  44.1709167&  0.047&  1&  208 & 290.4523750&  44.4416389&  0.056&  1  \\
149 & 290.1860000&  44.1786944&  0.049&  1&  209 & 290.3117500&  43.5235833&  0.063&  3  \\
150 & 290.5180833&  44.1755278&  0.053&  1&  210 & 289.6784167&  43.5988056&  0.055&  1  \\
151 & 290.1912500&  44.1814722&  0.057&  1&  211 & 290.6303333&  43.6313611&  0.059&  1  \\
152 & 290.2265417&  44.1813333&  0.057&  1&  212 & 290.2389583&  43.6634722&  0.046&  2  \\
153 & 290.1565000&  44.1824722&  0.048&  2&  213 & 290.5182917&  43.6611111&  0.059&  3  \\
154 & 290.3702083&  44.1818333&  0.053&  1&  214 & 289.7535833&  43.6744444&  0.049&  1  \\
155 & 290.5102083&  44.1826944&  0.043&  1&  215 & 290.2099583&  43.7167500&  0.047&  1  \\
156 & 289.9267917&  44.1885833&  0.041&  1&  216 & 289.7482500&  43.7306111&  0.054&  1  \\
157 & 290.5425833&  44.1863889&  0.042&  1&  217 & 290.7636667&  43.7708889&  0.042&  1  \\
158 & 290.7425833&  44.1866667&  0.053&  1&  218 & 290.2355417&  43.8809722&  0.052&  2  \\
159 & 290.2314167&  44.1972222&  0.054&  1&  219 & 290.5877083&  43.8909444&  0.047&  1  \\
160 & 290.2114167&  44.1983333&  0.049&  1&  220 & 290.8095833&  43.9025278&  0.057&  1  \\
161 & 289.8071250&  44.2018611&  0.047&  2&  221 & 290.1862500&  43.9295000&  0.059&  1  \\
162 & 289.6415417&  44.2065556&  0.051&  3&  222 & 290.1953333&  43.9551944&  0.063&  1  \\
163 & 289.9030417&  44.2097778&  0.049&  1&  223 & 290.4805000&  43.9631667&  0.057&  1  \\
164 & 290.3138750&  44.2164722&  0.049&  1&  224 & 289.8920000&  44.0131667&  0.056&  1  \\
165 & 290.5139583&  44.2146667&  0.054&  1&  225 & 290.7065000&  44.0601111&  0.054&  1  \\
166 & 290.1799583&  44.2181389&  0.065&  1&  226 & 290.2903750&  44.0733889&  0.049&  4  \\
167 & 290.0620000&  44.2213333&  0.053&  1&  227 & 290.2899583&  44.0948611&  0.048&  1  \\
168 & 290.3704167&  44.2205278&  0.061&  1&  228 & 290.8224583&  44.1462222&  0.044&  1  \\
169 & 290.3865833&  44.2237500&  0.053&  1&  229 & 290.4445833&  44.1853889&  0.066&  1  \\
170 & 290.7153750&  44.2231111&  0.044&  1&  230 & 289.7693333&  44.2120000&  0.058&  1  \\
171 & 289.7412500&  44.2336389&  0.050&  2&  231 & 290.8355417&  44.2953333&  0.062&  1  \\
172 & 290.6528333&  44.2307500&  0.050&  1&  232 & 290.5464167&  44.3239167&  0.049&  3  \\
173 & 290.6312083&  44.2315000&  0.045&  1&  233 & 289.7261250&  44.3862500&  0.059&  2  \\
174 & 289.9805833&  44.2438333&  0.048&  3&      &            &            &       &     \\
175 & 290.6877500&  44.2382500&  0.050&  1&      &            &            &       &     \\
176 & 290.1938750&  44.2438333&  0.053&  1&      &            &            &       &     \\
177 & 290.4701667&  44.2473333&  0.051&  1&      &            &            &       &     \\
178 & 290.4647917&  44.2484167&  0.056&  1&      &            &            &       &     \\
179 & 290.2862500&  44.2518611&  0.065&  1&      &            &            &       &     \\
180 & 290.7408333&  44.2532500&  0.054&  2&      &            &            &       &

\enddata
\vskip 1mm
\parbox{125mm}{Hubble type of the best-fit template $T$: 1=E, 2=S0, 3=Sa, 4=Sb, 5=Sc, 6=Sd, 7=Irr}

%\vskip -3mm \hskip 15mm \parbox{90mm} {$^a$ This column lists the seeing of the combined image.}
% \tablecomments{0.96\textwidth}{$^a$ This column lists the seeing of the combined image.}
% \tablenotetext{a}{This column lists the seeing of the combined image.}

\end{deluxetable}
 \def\baselinestretch{1.15}

\section{ANALYSES OF THE ENLARGED SAMPLE OF GALAXIES IN A2319}

\subsection{Spatial Distribution}

Fig.~9 shows the projected spatial distribution of sample II, superimposed with the contour map of surface density which is smoothed by a Gaussian window of $\sigma=1.'6$. The density contour deviates from spherical symmetry conspicuously, showing that
A2319 is far from a dynamically relaxed system with significant substructures. The central region of contour map appears to be elongated along the NW, and the outliers are elongated in the NE direction. The bright galaxies with known spectroscopic redshifts are mainly concentrated within a region with projected radius of ~20 arcmin. However, the newly-selected faint member galaxies are rather scattered within 1.0 Abell radius. Compared with Figure 3(a), the density around ~15 arcmin north has been enhanced, which makes the underlying substructure prominent. It is noticed that the surface density peak associated with A2319B moves substantially to the north. If it is not due to projection effect, the projected distance between two subclusters should be larger, which will lead to a decrease in the probability of gravitational binding.

%If a cluster is in the process of merging along the direction with a definite projection angle, say $\alpha > 30^{\circ}$ , with respect to the plane of sky, the substructures are likely to be detected by the localized variation in the velocity distribution (Colless \& Dunn 1996). To detect potential substructures, we apply the $\kappa$-test for sample II, and the result with $10^{3}$ simulations is also given in Table~3. For sample II, the probabilities $P(\kappa_n > \kappa^{\rm obs}_n)$ with various neighbor sizes ($5\leq n \leq11$) are slightly greater than those for sample I, but they are found to be less than 9 percent.

\vskip 4mm
\centerline{\framebox[9cm][c]{Fig.~9: Spatial distribution for sample II}}
\smallskip

 \subsection{ Color-Magnitude Relation for Early-Type Galaxies}

It is well known that brighter early-type galaxies in a cluster tend to be redder. This linear correlation between color and apparent magnitude for early-type galaxies is called the color-magnitude relation (hereafter C-M relation), which has been testified in some rich galaxy clusters (see Bower et al. 1992, and references therein). This correlation has been used for  selecting early-type member galaxies (Yuan et al. 2001, 2003; Liu et al. 2011 ).

Based on the morphology of the best-fit SED template given by the HYPERZ code, we obtain 118 early-type galaxies in sample I and 216 newly-selected early-types in sample II. Fig.~10 presents the correlation between the color index $b$-$h$ and $h$-band magnitude for all these early-type galaxies. A linear fitting is performed for the early-type galaxies in sample I (solid line): $b - h = -0.087(\pm0.036)h+3.62(\pm0.59) $, and the dashed lines denote the $\pm2\sigma$ deviation. It is clear that all the early-type galaxies follow a tight C-M relation within the range of $\pm2\sigma$ deviation, which demonstrates the reliability of our membership selection.

\vspace{4mm}
\centerline{\framebox[14cm][c]{Fig.~10: The color-magnitude relation for early-type galaxies in A2319 }}
\smallskip

\subsection{Environmental Effect on Star Formation Properties}

The star formation histories (SFHs) of member galaxies may shed some light on the evolution of their host cluster. With the evolutionary synthesis model, PEGASE (version 2.0, Fioc \& Rocca-Volmerange 1997, 1999), the star formation properties of the galaxies in A2319 are studied, assuming a Salpeter (1955) initial mass function (IMF) and a star formation rate (SFR) in exponentially decreasing form, $\rm{SFR}(t) \propto e^{-t/{\tau}}$, where $\tau$ is the time scale  ranged from 0.5 to 30.0\,Gyr. For avoiding the degeneracy between age and metallicity in the model, we adopt the same age of 12.97 Gyr for all member galaxies in A2319, corresponding to the age of first generation stars at $z=0.0557$. In our model, a zero initial metallicity of the interstellar medium (ISM) is taken. As a result, a series of rest-frame modeled spectra with various star formation histories is generated. After redshifted to the observer's frame and convolved with the transmission functions of the BATC filters, the library of template SEDs for the BATC multicolor photometric system (i.e., relative apparent magnitudes at 15 BATC filters) has been obtained. Based on the template SED library, we search for the best fit (in the $\chi^2$ sense) of the observed SEDs of 128 member galaxies with known spectroscopic redshifts. The star formation properties such as the SFR time scale $\tau$, mean ISM metallicity $\langle Z_{\rm ISM}\rangle$, and mean stellar age $\langle t_{\star}\rangle$ can be derived for each galaxy.

Fig.~11 shows the observed SEDs and the best-fitting templates for the BCG of A2319, CGCG~230-007, and a galaxy in A2319B, 2MASX~J19201995+4358346.  These two galaxies are different in SFR time scale and mean stellar age, and can be representatives of the galaxies in sample I. The uncertainties on the model parameters can be estimated by the Monte-Carlo simulation. For the galaxies with $16.0 < i < 17.0$ in sample I, the mean errors of magnitude for 15 filters are calculated as the measurement uncertainties on the observed BATC magnitude.  Then, 1000 random realizations of the BATC SEDs are generated by treating the simulated magnitude as a Gaussian random variable, on the basis of the best-fitting template magnitude perturbed by the typical uncertainties on the observed BATC magnitude.  For the BCG, we obtain a best-fitting model with $\tau_{\rm SFR}=1.00 \pm 0.30$ Gyr,  $\langle t_{\star}\rangle = 11.90 \pm 0.54$ Gyr, and $\langle Z_{\rm ISM}\rangle = 0.0356 \pm 0.0002$, which corresponds to the SFH of a typical elliptical galaxy. For the other galaxy,  2MASX~J19201995+4358346, we achieve a different SFH, with  $\tau_{\rm SFR}=2.80 \pm 0.25$ Gyr,  $\langle t_{\star}\rangle = 9.08 \pm 0.50$ Gyr, and $\langle Z_{\rm ISM}\rangle = 0.0350 \pm 0.0003$.

\vspace{4mm}
\centerline{\framebox[16cm][c]{Fig. 11:  The observed SEDs and the best-fitting templates of two representative galaxies. 
}}
\smallskip
%It should be noted that the ISM metallicity in our modeling is mainly constrained by the continuum shape, not by the emission line ratios. 

To study the environmental effect on star formation history, following Dressler (1980), we define the local surface density by nearest 10 neighboring galaxies in sample II, $\Sigma_{10}=10/\pi d^2_{10}$, where $d_{10}$ is the projected distance to the 10th nearest neighbor. Fig.~12 presents the star formation properties as the functions of two environment indicators, say, the clustercentric distance $R$ and local surface density $\Sigma_{10}$. The NED-given cluster center, ($19^h21^m08^s.8, +43^{\circ}57'30''$; J2000.0), is adopted to calculate the clustercentric distance. In the right panels of Fig.~12, the dashed line at $\Sigma_{10} = 350$ galaxies/Mpc$^{-2}$ separates the cluster coverage into dense and sparse regions.  This $\Sigma_{10}$ demarcation corresponds to a clustercentric radius of $R \sim 0.38$ Mpc, which separates the cluster coverage into core and outer regions in the left panels.  A comparison between the left and right panels shows that the star formation properties of the member galaxies in A2139 are found to be more dependent on  local density $\Sigma_{10}$, rather than on the clustercentric distance $R$. The panels (b), (d), and (e) clearly demonstrate a trend that the galaxies in the sparse regions are likely to be scattered in the domains of SFR time scale, stellar age, and ISM metallicity. On the other hand, the galaxies in the dense regions are found to have shorter SFR time scales, older stellar ages, and higher ISM metallicities, which is consistent with the morphology--density relation first pointed out by Dressler (1980). This trend can be well explained in the context of the hierarchical cosmological scenario (Poggianti 2004).

%It should be noted that the reliability of the ISM metallicity is not high because we assume a zero initial metallicity in our model. We know that the ISM metallicity can be well constrained by the emission line ratios, rather than by the SED shape. 

\vspace{4mm}
\centerline{\framebox[12cm][c]{Fig. 12:  Star formation properties for the 128 galaxies  in A2319
%as functions of cluster centric distance $R$ and the local surface density $\Sigma_{10}$
%The star formation properties include the SFR time scale $\tau$, metallicity, and the mean stellar ages weighted by mass and light.
}}
\smallskip

As shown in the panels~(c) and (d), some outlier member galaxies in the sparse regions are likely to possess younger stellar populations, resulting in a smaller mean stellar age weighted by mass. Panels (e) and (f) show that the outlier galaxies have a greater probability of having a lower mean ISM metallicity. It is considered that the galaxies in the dense core region tend to be more massive and luminous. The underlying physical correlation is the luminosity--metallicity relation (Melbourne \& Salzer 2002) and the mass--metallicity relation (Tremonti et al. 2004). This idea is that more massive galaxies form fractionally more stars in a Hubble time than their low-mass counterparts, and  metals are selectively lost from faint galaxies with shallow potential wells via galactic winds.

It should be noted that  the clustercentric distance $R$ is a good environmental indicator for a well-relaxed cluster of galaxies (e.g., A2589, Liu et al. 2011). For the galaxies in A2319, the local surface density $\Sigma_{10}$ proves to be a better environmental indicator, indicating that A2319 is a dynamically complex cluster. An alternative explanation is that star formation activities of galaxies in a cluster with ongoing merger events might be more sensitive to galaxy-scale gravitational interaction, rather than to the cluster-scale environment.

May the cluster merger affect the star formation histories of the resident galaxies?  Lavery \& Henry (1994) proposed that the star formation can be triggered by galaxy-galaxy interactions in clusters with inter-mediate redshifts. Numerical simulations by Gnedin (1999) showed that a time-varying cluster potential may cause a sequence of strong tidal shocks on individual galaxies. The shocks would probably take place over a wide region of the cluster and enhance the galaxy-galaxy merger rates. We once presented the star formation properties of 68 member galaxies in nearby (at $z=0.0414$) relaxed cluster A2589 (Liu et al. 2011). It should be interesting to compare the star-forming properties for these two clusters in very different dynamical states. However, due to different completenesses of spectroscopy and lack of spectral analysis, it is difficult to quantitatively compare the star-forming properties of these two clusters. 

A statistics about the fraction of galaxies with evidence for current star formation can be done. It is well known that the time scale of star formation may reflect current star formation activity. The Sb galaxies have a typical time scale of $\tau_{\rm SFR} = 5$ Gyr  (see HyperZ manual, Bolzonella et al. 2000).  The fraction of galaxies with $\tau_{\rm SFR} > 5$ Gyr in A2589 is about $3/68=4.4\%$, while this fraction in A2319 is about  $10/128=7.8\%$. All these 10 galaxies locate at the sparse region of A2319.  For a higher threshold of time scale, $\tau_{\rm SFR} = 15$ Gyr, which corresponds to the typical Sc galaxies, no such galaxies in A2589 are found, while there are four star-forming galaxies in A2319.  This may suggest that the galaxy-scale turbulence stimulated by the merging of subclusters might have played an important role in triggering the star formation activity. 

%It should be noted that the reliability of the ISM metallicity is not high because we assume a zero initial metallicity in our model. We know that the ISM metallicity can be well constrained with the measurement of emission line ratios, rather than the SED shape. 

\section{SUMMARY}

Previous optical, radio and X-ray studies showed that A2319 is not a regular and dynamically relaxed galaxy cluster. This paper presents our multicolor optical photometry for this cluster, on the basis of 15 intermediate filters in the BATC system that cover almost the whole optical wavelength domain. The SEDs of more than 30,000 sources down to $V=20^{m}$ in our field of view ($58'\times58'$) have been obtained. There are 142 normal galaxies with known spectroscopic redshifts in our field, among which 128 galaxies are selected as member galaxies of A2319 (i.e., sample I), by using the "shifting-gapper" method. Considering a very large dispersion in the rest-frame velocity, $1622^{+91}_{-70}$ \kms, and a cold front in the Chandra image, we suggest a merger model where the trajectory lies approximately $50^{\circ}\pm20^{\circ}$ out of the plane of the sky.  The contour map of projected density and localized velocity structure for sample I confirm the substructure so-called  A2319B, at $\sim10'$ NW with respect to the main concentration A2319A. A probable substructure A2319C is detected at $\sim 30'$ NW. Then, the KMM algorithm assigns 106 and 22 galaxies to A2319A and A2319B, respectively, and these two subclusters are probably gravitationally bound.

Three color-color diagrams are taken to make star-galaxy separation. The software SExtractor is applied to the $u$-band  image with better seeing and resolution which is obtained by the Bok 2.3m telescope, and further cross-identification results in 927 galaxies  brighter than $h_{\rm BATC}$ = 19.0. Based on the photometric redshift estimate of these galaxies, 233 member galaxies are newly selected as member candidates in A2319, after an exclusion of false candidates with contaminated BATC SEDs by one-by-one eyeballing check. Based on the enlarged sample (i.e., sample II), spatial distribution of A2319 are investigated. The substructure A2319B appears prominent, and its centroid moves substantially to the north.  All early-type galaxies in sample II are found to follow a tight C-M relation.

Assuming a Salpeter IMF and an exponentially decreasing SFR, the star formation properties of the galaxies in A2319 are derived with the evolutionary synthesis model, PEGASE. A strong environmental effect on the star formation histories has been found in the manner that the galaxies in the sparse regions are likely to be scattered in the domains of SFR time scale, stellar age, and ISM metallicity. On the other hand, the galaxies in the dense regions seem to have shorter SFR time scales, older stellar ages, and higher ISM metallicities. In A2319, the local surface density is the driving factor of star formation histories, rather than the clustercentric distance, indicating that A2319 is a dynamically complex cluster.

Compared with the well-relaxed cluster A2589, a statistics about the fraction of star-forming galaxies suggest that the galaxy-scale turbulence stimulated by the subcluster merger might have played an important role in triggering the star formation activity.  For a further investigation on the merging dynamics and star formation history in A2319, it is necessary to take a homogeneous deep spectroscopic observations for at least 400 member galaxies. The detailed spectral analysis of such a large sample may shed some light on how the merging dynamics in the cluster scale affects the star formation in member galaxies, and on how and when the cluster-scale radio halo was evoked.

%\vskip 1cm

\section*{Acknowledgements}

We thank the anonymous referee for his/her thorough reading of this paper and many invaluable suggestions. This work is funded by the National Natural Science Foundation of China (NSFC) (Nos. 11173016, 11273006, 10873016) and by the National Basic Research Program of China (973 Program; grant No. 2007CB815403). This research has made use of the NASA/IPAC Extragalactic Database (NED), which is operated by the Jet Propulsion Laboratory, California Institute of Technology, under contract with the National Aeronautics and Space Administration. We would like to thank Dr. Zhaoji Jiang, Jun Ma, Zhou Fan, Hu Zou, and Zhenyu Wu at National Astronomical Obseratories of China for their valuable discussions. We also appreciate the assistants who contributed their hard work to the observations.

%\clearpage

%% -----------------------------------------------------------

\clearpage

%fig1
\begin{figure}
\epsscale{0.5}
\plotone{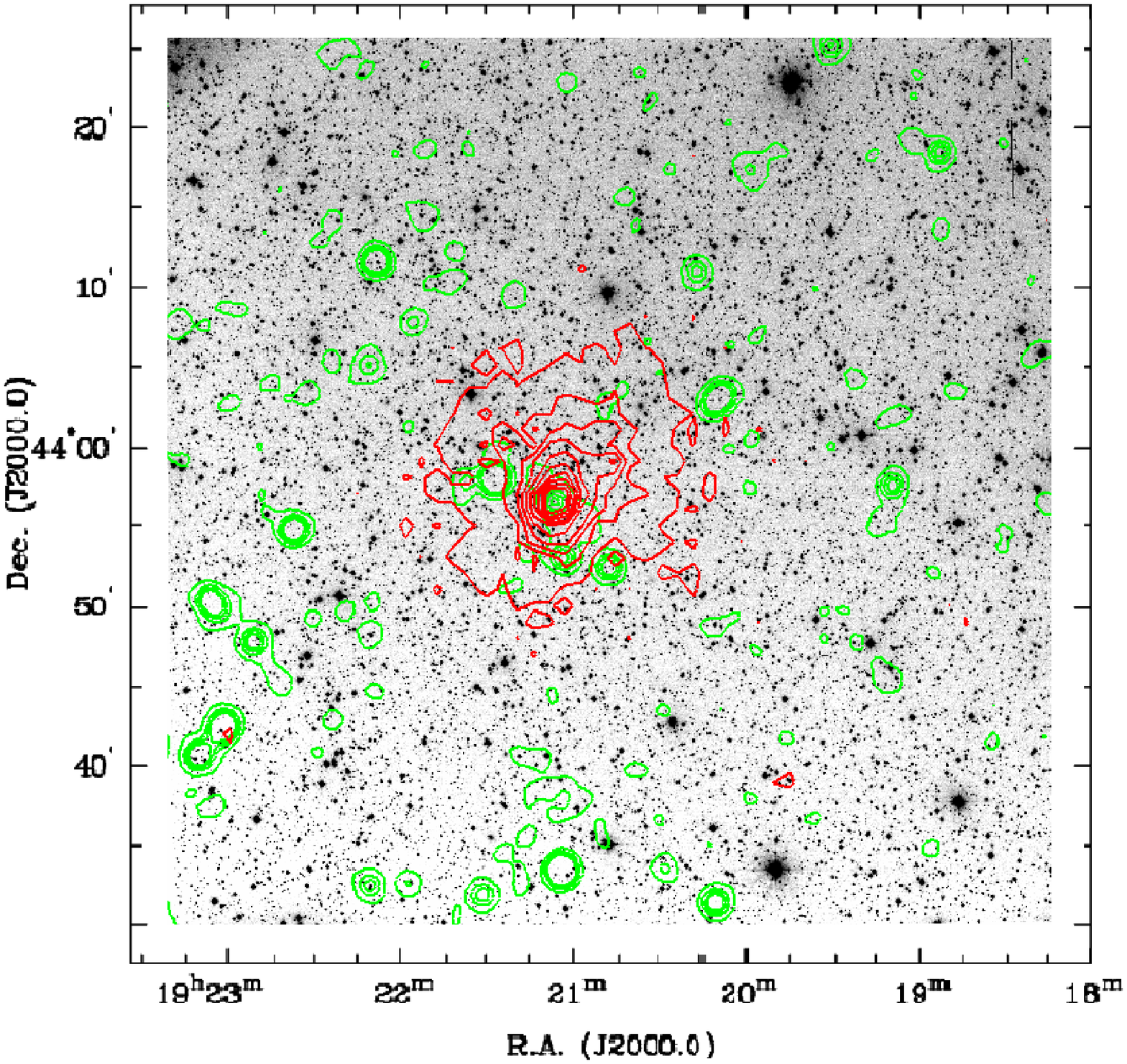}
\caption{\small Smoothed counters of the ROSAT image (0.1-2.4 keV) (red line) and the NVSS map at 1.4 GHz (green line), superimposed on the BATC-$d$ band image. The sizes of both Gaussian smoothing windows are 1 arcmin.}
\end{figure}

%\clearpage

%fig2
\begin{figure}
\epsscale{0.95}
\plottwo{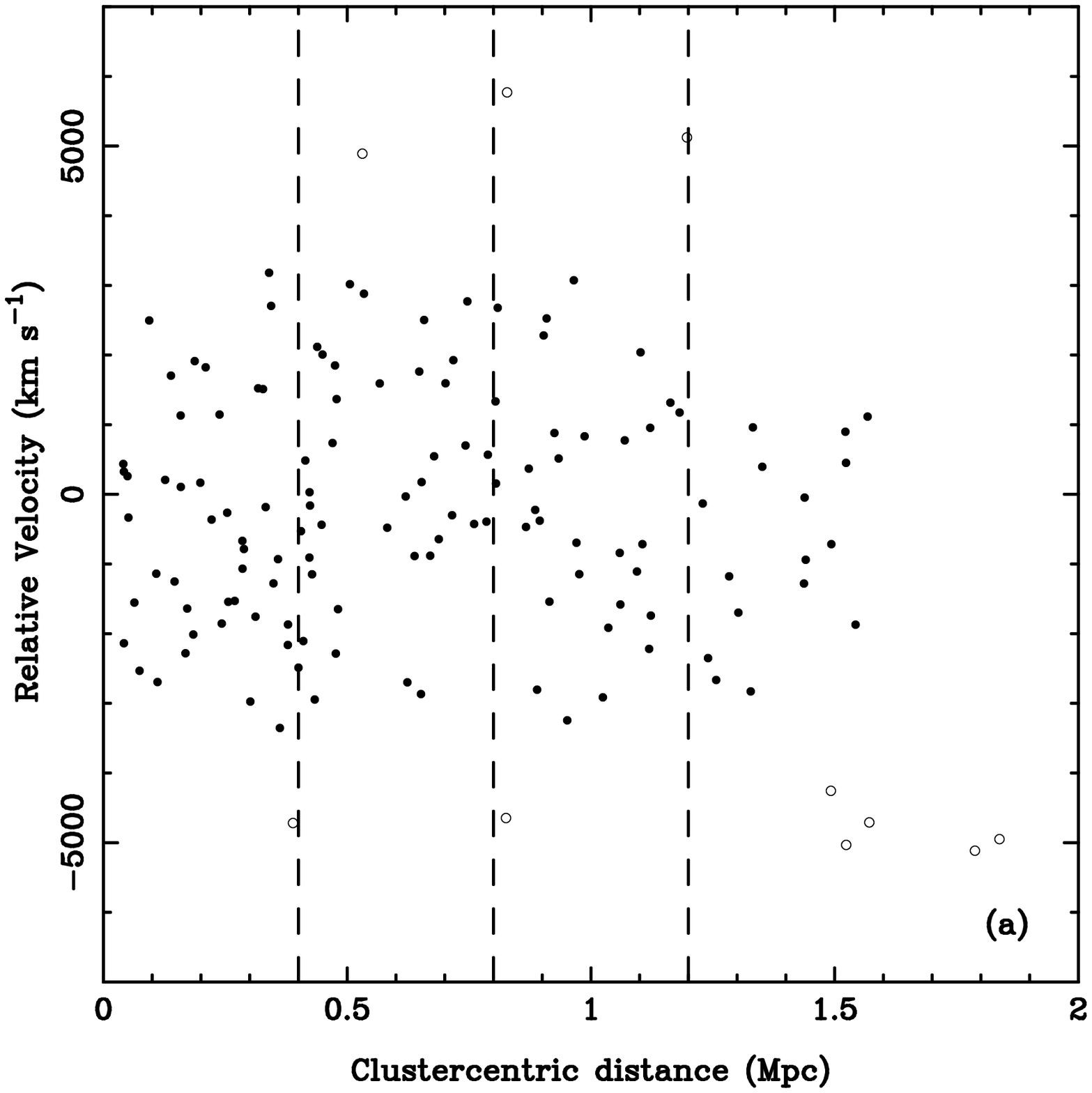}{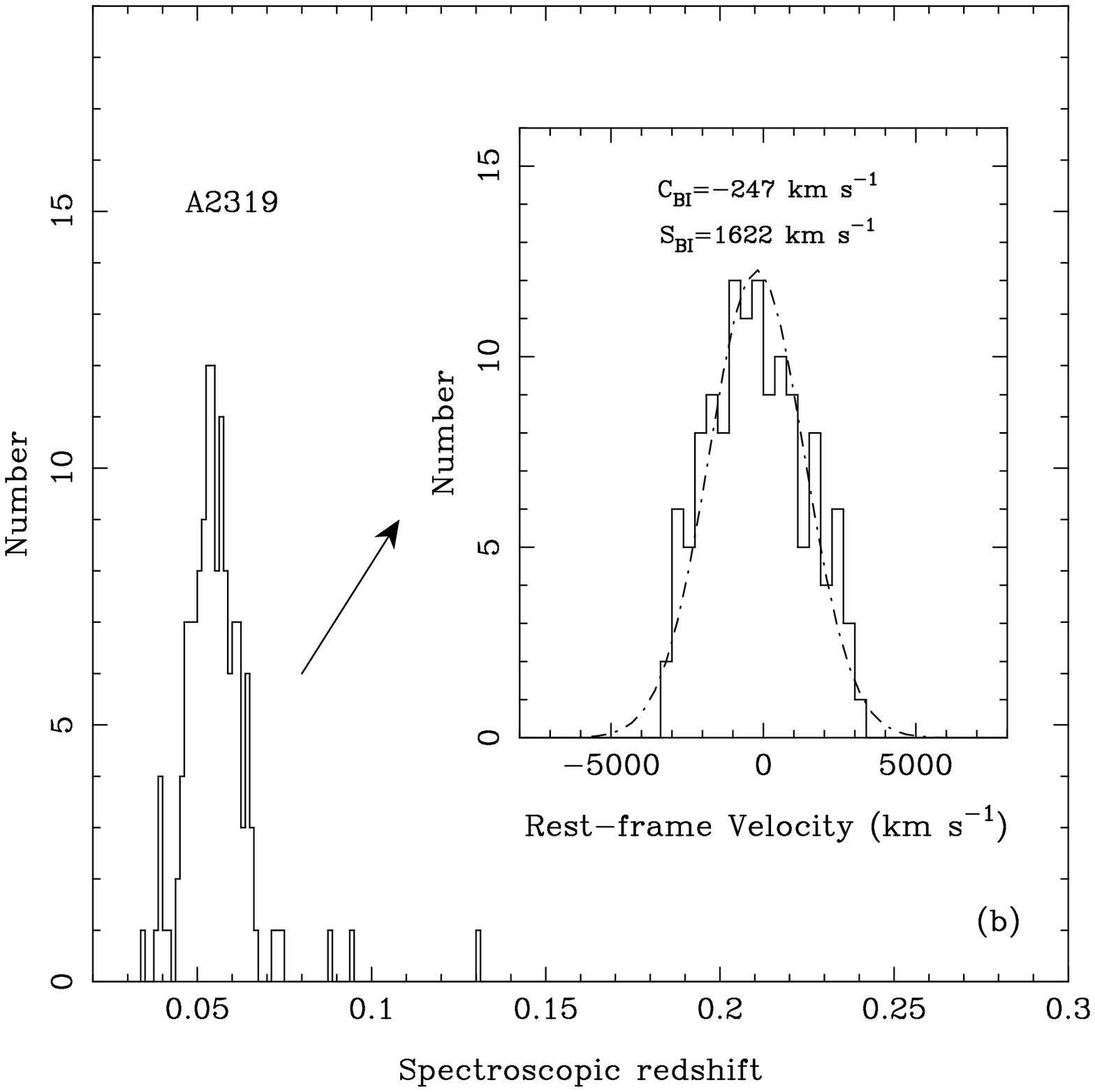}
\caption{\small (a) Relative velocities for the 138 galaxies with $0.033<z_{sp}<0.075$ as the functions of clustercentric distances. The open circles indicate the interlopers rejected by the ``shifting-gapper'' method. Dashed lines denote four bins of clustercentric distance. (b) Distribution of spectroscopic redshifts for 142 galaxies in the A2319 field. The embedded panel shows the histogram of rest-frame velocities for 128 member galaxies.}
\end{figure}

%\clearpage

%fig3
\begin{figure}
\epsscale{0.95}
\plottwo{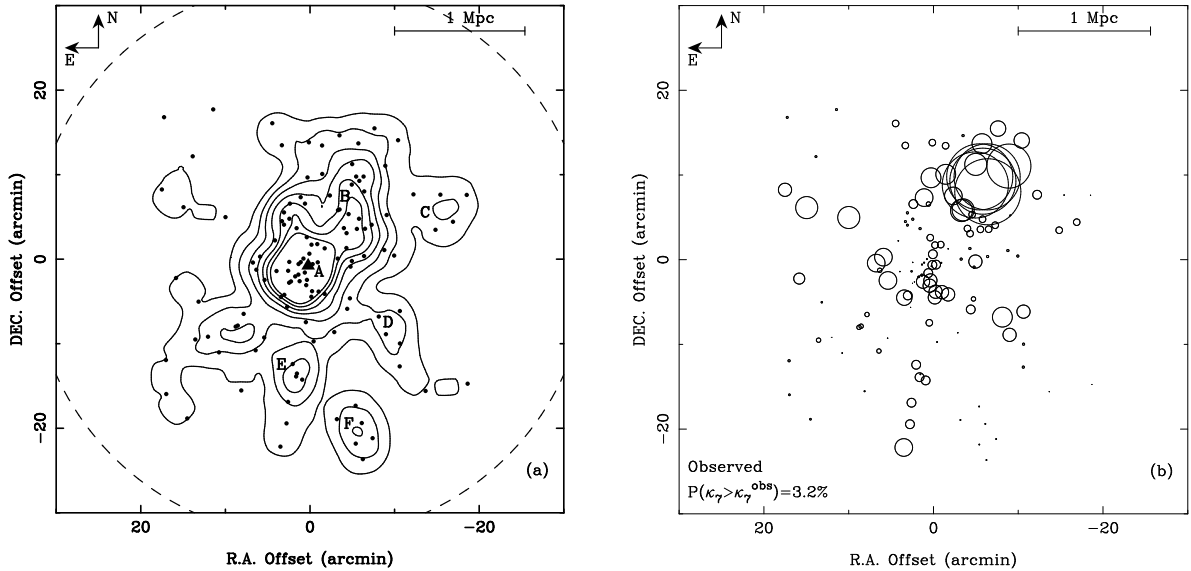}{fig03b.eps}
\caption{\small (a) Spatial distribution for 128 member galaxies in sample I. The contour map of the surface density is smoothed by a Gaussian window of $\sigma=1.'6$. The contour levels are 0.04, 0.09, 0.14, 0.19, 0.24, 0.29 and 0.34 arcmin$^{-2}$, respectively. The dashed circle shows a typical region of rich clusters with a radius of 1.5 $h^{-1}$Mpc from the cluster center. The brightest cluster galaxy (BCG), namely, CGCG~230-007, is denoted by the filled triangle. (b) Bubble plot showing the localized variation in velocity distribution with seven nearest neighbors for sample I.}
\end{figure}

%\vskip 1cm
%\clearpage

%fig4
\begin{figure}
\epsscale{1.0}
\plottwo{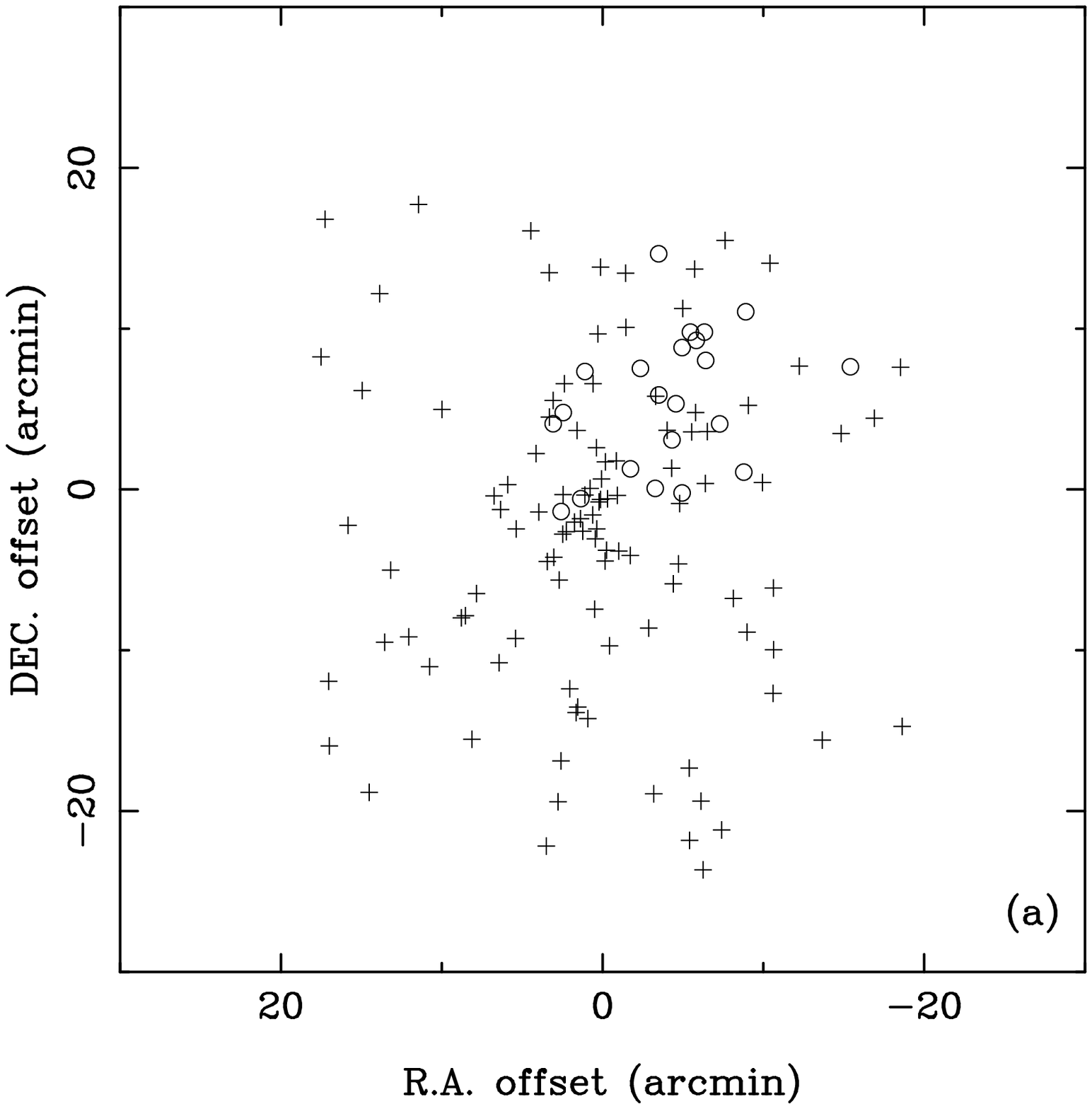}{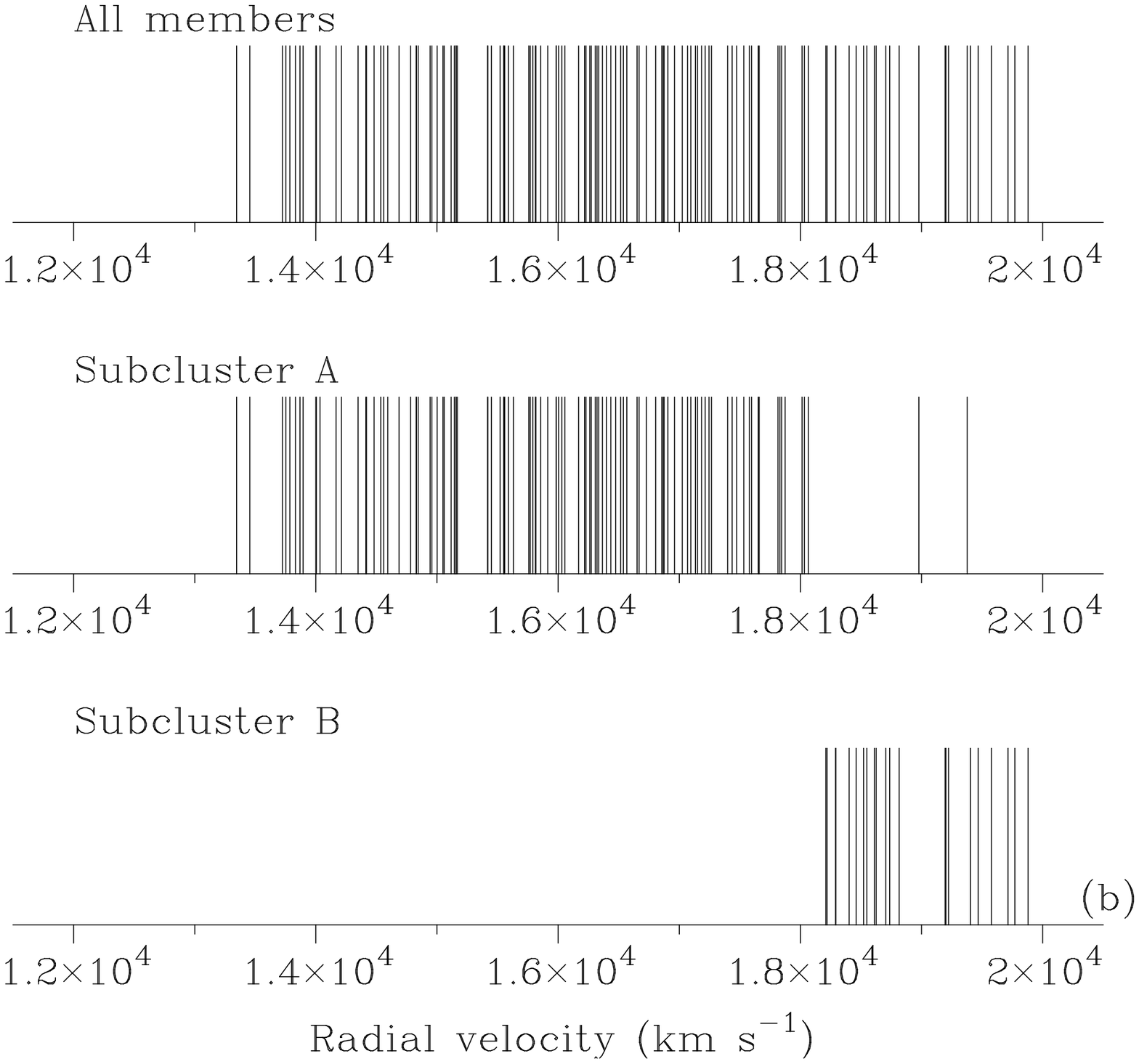}
\caption{\small (a) Spatial distribution of the spectroscopically confirmed member galaxies in subcluster A (denoted by ``+'') and subcluster B (denoted by the open circles). (b) Stripe density plot of radial velocities for sample I.}
\end{figure}

%fig5
\begin{figure}
\epsscale{0.5}
\plotone{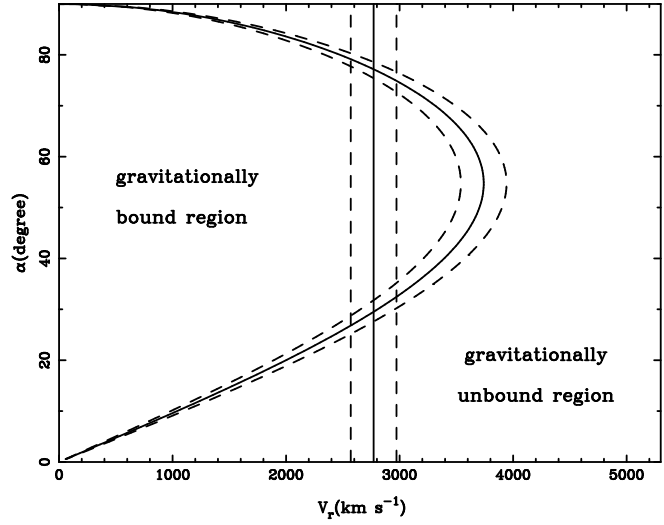}
\caption{\small Bound and unbound orbits as a function of $\alpha$,
the projection angle of A2319A and A2319B, and $V_{r}$,
their relative radial velocities. The dashed curves represent the uncertainty of the Newtonian binding criterion when the uncertainty of total virial mass is taken into account. The vertical lines represent the relative radial velocity,
$V_{r}= 2771 \pm 202 $ km s$^{-1}$, between the two clusters in the rest frame of A2319.}
\end{figure}

%fig6a,b,c
\begin{figure}
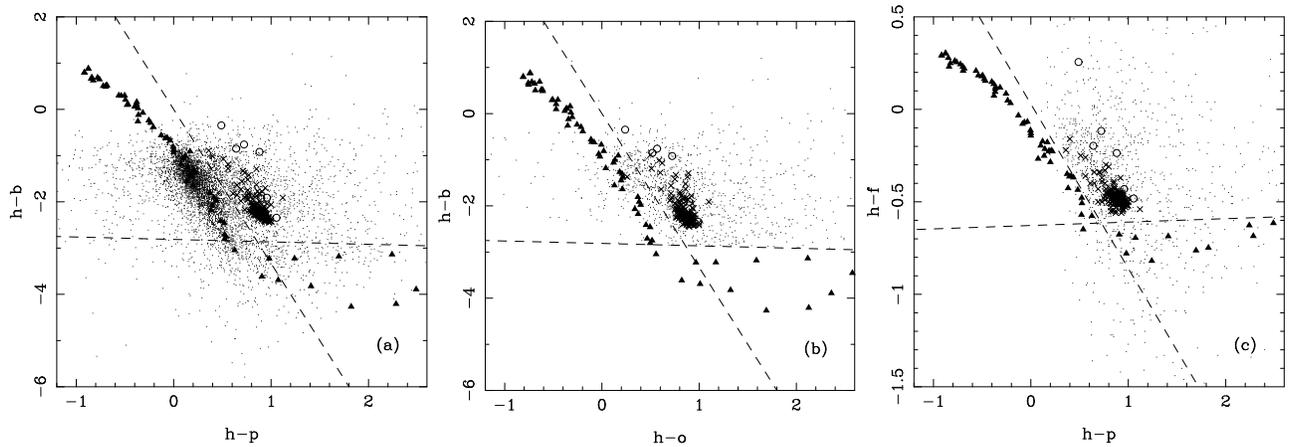

\epsscale{0.32}
\plotone{fig06a.eps}
\plotone{fig06b.eps}
\plotone{fig06c.eps}
\caption{\small The color-color diagrams for star-galaxy separation, including all types of stars in the SED template library ({\em triangles}), various Hubble type of galaxies with template SEDs ({\em open circles}), the spectroscopically confirmed galaxies ({\em crosses}) and the BATC detected sources ({\em small dots}). The dashed lines can be taken as boundaries of separation.}
\end{figure}

%fig.7
\begin{figure}
\epsscale{0.8}
\plotone{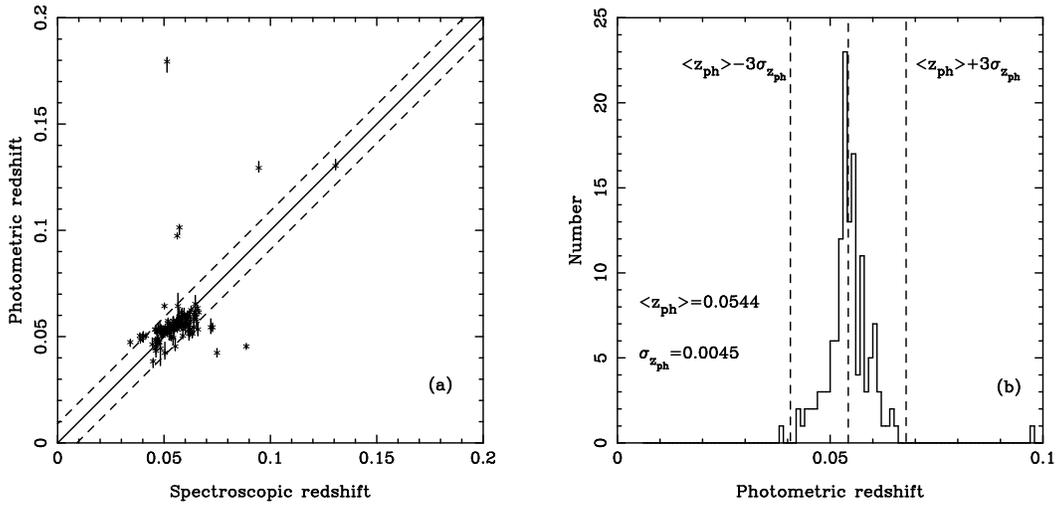}
\caption{\small (a) The $z_{\rm ph} - z_{\rm sp}$ comparison for the 142 bright galaxies with known $z_{\rm sp}$ values in the region of A2319. The solid line corresponds $z_{\rm ph}=z_{\rm sp}$,  and the dashed lines denote 2$\sigma$ deviation of the photometric redshift.
(b) The $z_{\rm ph}$ distribution for these galaxies. The dashed lines represent the photometric redshift range of selection criterion.}
\end{figure}

%fig.8
\begin{figure}
\epsscale{0.65}
\plotone{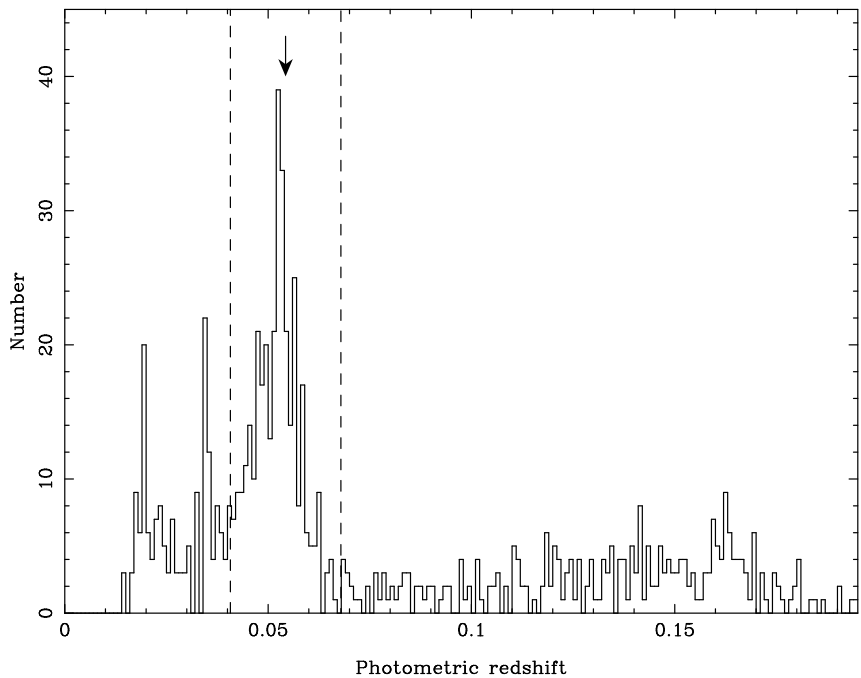}
\caption{\small The $z_{\rm ph}$ distribution for the 927 galaxies. The selection criterion is shown by the arrow and dashed lines, which denote the $z_{ph}$ mean and $\pm 3\sigma$, respectively.}
\end{figure}

%fig.9
\begin{figure}
\begin{center}
\epsscale{0.5}
\plotone{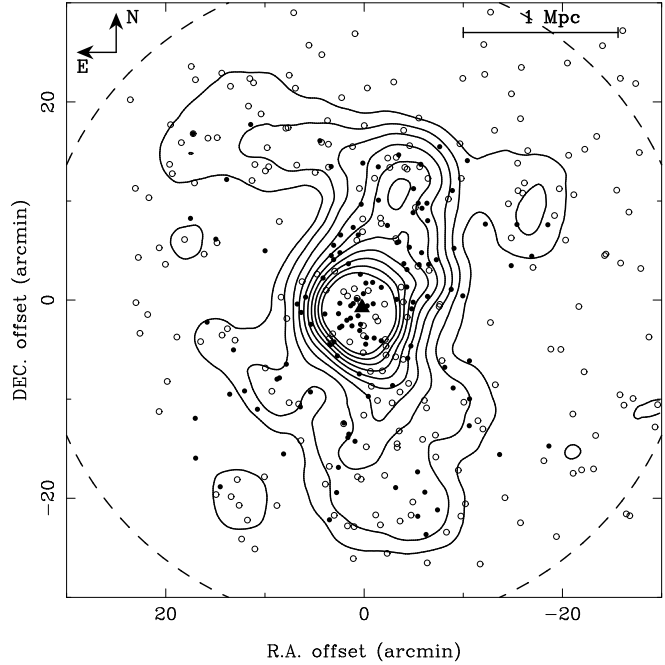}%{fig09b.eps}
\caption{\small Spatial distribution of the galaxies in sample II, superposed with the contour maps of surface density which has smoothed by a Gaussian window of $\sigma=1.'6$. The contour levels are 0.12, 0.17, 0.22, 0.27, 0.32, 0.42, 0.47, 0.52, and 0.57 arcmin$^{-2}$. }
\end{center}
\end{figure}

%fig. 10
\begin{figure}
\epsscale{0.5}
\plotone{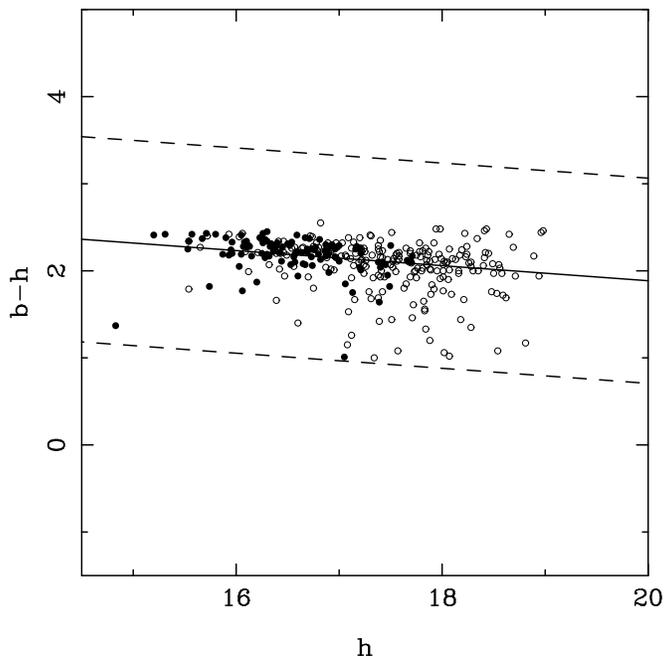}
\caption{\small Color-magnitude correlation for all the early-type galaxies in A2319, including the early-types in sample I ({\em filled circles}) and newly-selected early-type member candidates ({\em open circles}). A linear fit for the early-type galaxies in sample I is shown by solid line, and dashed lines correspond to $2\sigma$ deviation. }
\end{figure}

%fig.11
\begin{figure}
\begin{center}
\epsscale{0.55}
\plotone{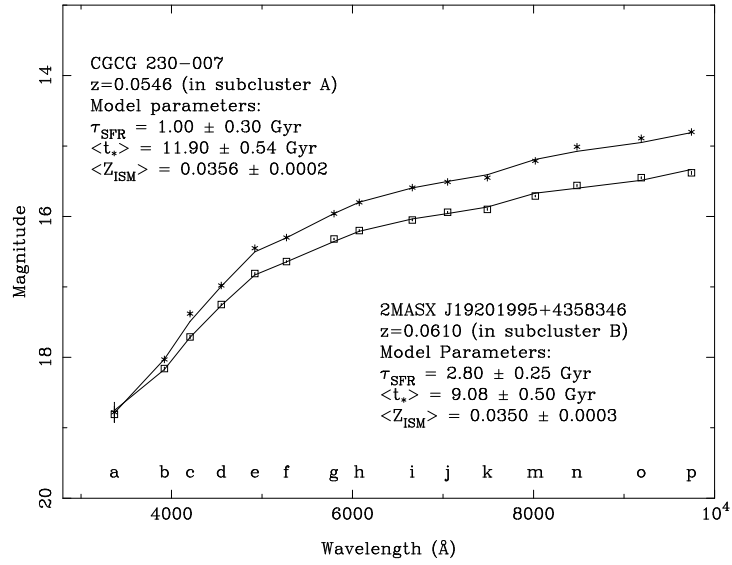}
\caption{The observed SEDs and the best-fitting templates for the BCG, CGCG~230-007, and a typical galaxies in A2319B, 2MASX~J19201995+4358346. The model parameters are also given in the plot. The uncertainties of model parameters are estimated by the Monte-Carlo  simulations where random realizations of the BATC magnitudes are measured from the best-fitting template and perturbed by the measurement uncertainties on the observed BATC magnitudes.}
%\label{fig11}
\end{center}
\end{figure}

%fig.12
\begin{figure}
\begin{center}
\epsscale{0.9}
\plotone{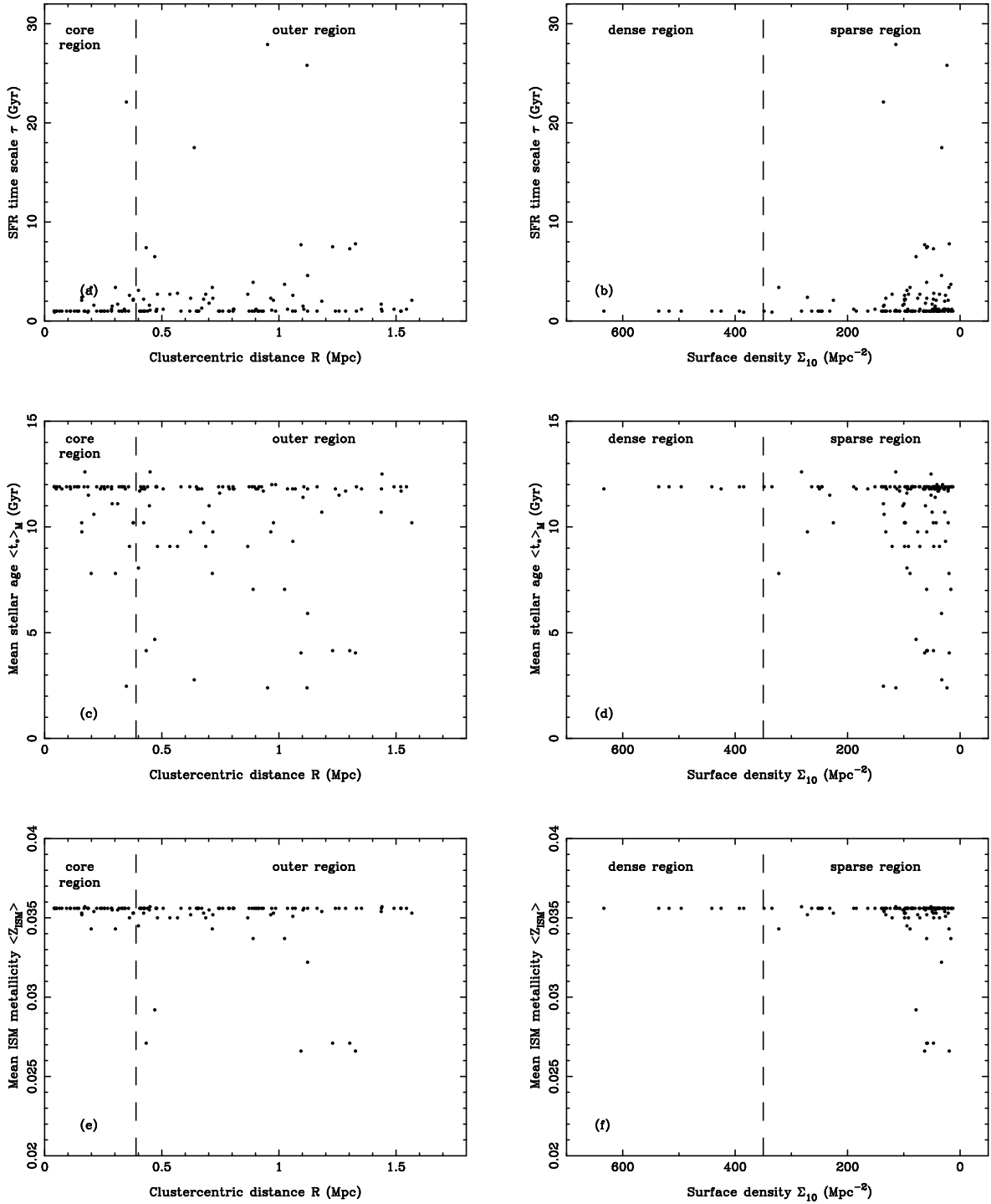}
\caption{The SFR time scales ($\tau$),  mass-weighted mean stellar ages ($\langle t_* \rangle_{\rm M}$), and mean ISM metallicities ($\langle Z_{\rm ISM} \rangle$) for the 128 galaxies with known redshifts in A2319, as the functions of clustercentric distance $R$ and local surface density $\Sigma_{10}$. }
%\label{fig12}
\end{center}
\end{figure}

%% If you are not including electronic art with your submission, you may
%% mark up your captions using the \figcaption command. See the
%% No more than seven \figcaption commands are allowed per page,
%% so if you have more than seven captions, insert a \clearpage

\clearpage
\def\baselinestretch{1.2}

\clearpage
%\input{tab2.tex}
%\begin{figure}
%\epsscale{1.0}
%\plotone{tab2_1.eps}
%\end{figure}
 \def\baselinestretch{0.85}

\clearpage
%\input{tab4.tex}
%\begin{figure}  \epsscale{1.04}
%\plotone{tab4_1.eps}
%\end{figure}  \clearpage
%\begin{figure} \epsscale{1.04}
%\plotone{tab4_2.eps}
%\end{figure}  \clearpage
%\begin{figure}  \epsscale{1.04}
%\plotone{tab4_3.eps}
%\end{figure}

\end{document}